# Throughput Optimal Switching in Multi-channel WLANs

Qingsi Wang and Mingyan Liu
Department of Electrical Engineering and Computer Science,
University of Michigan, Ann Arbor, MI 48109

***Abstract*—We observe that in a multi-channel wireless system, an opportunistic channel/spectrum access scheme that solely focuses on channel quality sensing measured by received SNR may induce users to use channels that, while providing better signals, are more congested. Ultimately the notion of channel quality should include both the signal quality and the level of congestion, and a good multi-channel access scheme should take both into account in deciding which channel to use and when. Motivated by this, we focus on the congestion aspect and examine what type of dynamic channel switching schemes may result in the best system throughput performance. Specifically we derive the stability region of a multi-user multi-channel WLAN system and determine the throughput optimal channel switching scheme within a certain class of schemes.**

## I. Introduction

Advances in software defined radio in recent years have motivated numerous studies on building agile, channel-aware transceivers that are capable of sensing instantaneous channel quality [1], [2], [3]. With this opportunity comes the challenge of making effective opportunistic channel access and transmission scheduling decisions, as well as designing supporting system architectures. There have been extensive studies on dynamic channel access in a multi-user, multi-channel wireless system, see e.g., [4], [5]. By allowing users to dynamically select which channel to use for transmission, these schemes aim to improve the system performance, typically measured by the total (or per user) throughput, the average packet delay and etc, compared to a system with a single channel or more static channel allocations. The main reason behind such improvement lies in temporal, spatial and spectral diversity. That is, the quality of a channel perceived by a user is time-varying, user-dependent, and channel-dependent.

Within this context we make the additional observation that there is also a *congestion diversity* in that a channel with fewer number of competing users presents better quality for a user. This is particularly true in a random access setting, where a large number of competing users can induce large backoff timer values that in turn lead to longer waiting time and lower throughput. We note that in a multi-channel system, an opportunistic access scheme that solely focuses on channel quality sensing as a result of random fading and shadowing, e.g., by measuring received SNR [6], [5], may induce users to use channels that, while providing better signals, are more congested. This can reduce the expected performance gain, or even turn gain to loss. Ultimately the notion of "channel quality" should include both the signal quality and the level of congestion, and a good multi-channel access scheme should take both into account in deciding which channel to use and when.

Motivated by the above, in this study we examine the possibility of utilizing congestion diversity to promote certain performance measures, e.g., throughput. As mentioned above, our ultimate goal is to construct an opportunistic channel access scheme that is aware of both signal quality and user congestion. However, in the present paper we will limit our attention to addressing the congestion aspect only; a good understanding of this aspect is a crucial first step in this effort.

Specifically, we ask the question of what type of dynamic channel switching schemes will give the best performance in a multi-channel WLAN. This will be evaluated using the notion of stability region of a scheme. This is because more effective resource allocation and sharing can achieve a lower overall congestion level, thus expanding the range of sustainable arrival rates and resulting in a larger stability region. The scheme with the largest such region is commonly known as the throughput optimal scheme. With this objective, we set out to study the stability region of a multi-channel WLAN system where users are allowed to dynamically switch between channels.

The main contributions of this paper are as follows.

- A mean-field based model is constructed to characterize the stability region of a multi-channel WLAN system. We show that the size of the backoff window plays a decisive role in shaping the corresponding stability region: when the backoff window is sufficiently large, the stability region is convex; as the window size decreases it evolves into a concave region.
- Using this mean-field model, we provide an analytical justification of using channel-switching policies that achieve load balance in systems with symmetric channels. This is then extended to systems with asymmetric channels.
- We propose several simple heuristic implementations of the channel-switching policies presented in this paper.

802.11 DCF has been very extensively studied in the literature, ranging from throughput performance in the saturated regime [7], [8] and the non-saturated regime [9], [10], to its rate region [11], [12], to name a few. To the best of our knowledge, however, none has studied multi-channel WLAN in the context of stability region. Works most relevant to ours

This work is supported by NSF grant CIF-0910765, ARO grant W911NF-11-1-0532, and NASA grant NNX09AE91G.

include ones on the stability region of slotted Aloha (e.g., [13]) and the rate region of 802.11 DCF [11], [12].

In the remainder of the paper, we first introduce a system of equations to characterize the stability region of a single channel WLAN consisting of multiple users within a single interference domain (Section III) followed by numerical results (Section IV). We then extend the same method to characterize the stability region of a multi-channel system and use this result to determine the throughput optimal channel switching schemes within a class (Section V). We also discuss how such schemes may be implemented in practice (Section VI).

## II. System Model and Preliminaries

Consider a multiple access system using the IEEE 802.11 DCF. There are $N$ nodes (or users interchangeably), indexed by the set $\mathcal{N} = \{1, 2, \ldots, N\}$, each with an infinite buffer, one transceiver and uses the same parameterization. We assume the channel is ideal and there is no MAC-level packet discard, i.e., there is no retransmission limit of a packet after collision. Throughout the analysis we also adopt a few other simplifying assumptions to make the problem tractable; these will be stated in the context to which they apply. It should be noted that due to the complexity of the problem, successive simplification in the modeling effort is a rather common practice and has been used in most if not all previous works. We later show that these simplifications do not impact the accuracy of the model under normal operating parameter values.

The key to our method is to model the queue at each node with a service process defined by 802.11 DCF as a *slotted mean field Markov chain* [14].

*Definition 1:* A *virtual backoff timer* of the system (or of a virtual node) is a universal timer for all nodes in the system: it counts down indefinitely, alternating between the count-down mode (when nodes in the system are counting down) and the freezing mode (when some node in the system is transmitting). The slot time is thus a random variable.

*Remark 1:* The above definition provides a universal slot time for all nodes in the system, and we will assume that the real backoff timer at each node is synchronized to this virtual timer on slot boundaries. The motivation behind such a construction originates from the principal difficulty in modeling a non-saturated system: the service process at each node runs in embedded time in terms of a slot, which is a random variable, whereas the packet arrival process is more naturally described in real time [14]. This difficulty does not exist in saturated analysis, see e.g., [7], where arrival processes do not play a role.

We next introduce three key assumptions in our model, followed by a discussion on their implications and limitations.

- **(A1)** The MAC layer arrival process at node $i$ is Poisson with rate $\lambda_i$ bits per second.
- **(A2)** The service time of a packet, i.e. the time from the initial backoff to successful transmission, is (i) exponential with service rate $\mu_i$ at node $i$, and (ii) independent of all arrival processes.
- **(A3)** Let $S(t)$ be the counting process of the number of slots accumulated up to time $t$. $S(t)$ is assumed to be (i) independent of $Q_i(t)$, and (ii) renewal.

Denote by $\boldsymbol{\lambda} = (\lambda_1, \lambda_2, \ldots, \lambda_N)$, and by $\{Q_i(t)\}_t$ the queueing process at node $i$ (also written as $Q_i(t)$ for simplicity), i.e., the number of packets in node $i$'s MAC queue at time $t$. We now formally define the stability region of system as follows.

*Definition 2:* The *stability region* $\Lambda$ is the set of all $\boldsymbol{\lambda} \in \mathbb{R}_+^N$ such that $Q_i(t)$ admits a stationary distribution for all $i$ with arrival rates $\boldsymbol{\lambda}$ under the 802.11 DCF scheme.

The above simplifying assumptions are not entirely realistic. Typically, due to congestion control by upper-layer protocols, e.g., TCP, the arrival process to the MAC layer is neither Poisson nor independent of the service process. However, as our objective is to explore the inherent properties of 802.11 DCF, the independence assumption is adopted to decouple the MAC layer from upper layers, while the Poisson and exponential assumptions are adopted to avoid technicalities that can obscure the main insight. Note that under the mean field methodology, each node is analyzed in isolation from the activities of all other nodes which are collectively regarded as an aggregate stationary process. Within such a framework the packet service time is taken to be stationary (see e.g., Bianchi's well-known mean field Markovian model of the service process [7]).

With **A1** and **A2**, $Q_i(t)$ is then a well-defined $M/M/1$ queue, and for a given $\boldsymbol{\lambda}$, $\boldsymbol{\lambda} \in \Lambda$ if and only if $Q_i(t)$ is positive recurrent. Equivalently we may consider the utilization factor $\rho_i$ at node $i$, given by $\rho_i = \min\{\frac{\lambda_i}{\mu_i}, 1\}$: the queue is stable if and only if $\rho_i < 1$. If $Q_i(t)$ is positive recurrent, then it is ergodic and we have $\lim_{t\to\infty} P(Q_i(t) > 0) = 1 - \pi_i(0) = \rho_i$, where $\pi_i$ is the stationary distribution of $Q_i(t)$. If $Q_i(t)$ is transient or null recurrent, in which case $\rho_i = 1$, we have $\lim_{t\to\infty} P(Q_i(t) = 0) = 0 = 1 - \rho_i$. Therefore, $\rho_i$ is asymptotically given by $\lim_{t\to\infty} P(Q_i(t) > 0)$ in all cases in our model.

For technical reasons we will also consider the *embedded* queueing process $\hat{Q}_i(n)$, $n = 1, 2, \cdots$, defined as $\hat{Q}_i(n) := Q_i(T_n)$, where $T_n$ is the time of the $n$th slot boundary. $\hat{Q}_i(n)$ is thus a discrete-time process constructed by observing $Q_i(t)$ at slot boundaries.

For an arbitrary process $S(t)$, $\hat{Q}_i(n)$ is not necessarily Markovian. However, given assumption **A3**, durations between slot boundaries are i.i.d., constituting sampling periods that are independent of $Q_i(t)$. Hence $\hat{Q}_i(t)$ is a discrete-time Markov chain under our assumption[1]. It's worth noting that **A3** does not exactly hold in reality because the slot length is a function of a node's activity, and thus the state of its queue, even with the mean field simplification of other nodes' behavior (this is more precisely shown in the appendix). However, this dependence weakens when the number of nodes or the backoff window size is sufficiently large. We empirically show that this assumption does not impact the accuracy of prediction even with a small node population and backoff window size.

[1]This claim is proved in Appendix G.

Let $\hat{\rho}_i$ denote the utilization factor under the discrete-time system $\hat{Q}_i(n)$. In general $\hat{\rho}_i \neq \rho_i$. Indeed we show in Appendix A that $\hat{\rho}_i \leq \rho_i$ where equality holds if and only if $\rho_i = 1$ or $\rho_i = 0$, i.e., node $i$ is either saturated or idle. Similar to $\rho_i$, $\hat{\rho}_i$ is asymptotically given by $\lim_{n\to\infty} P(\hat{Q}_i(n) > 0)$.

We will adopt Bianchi's decoupling approximation [7] as another key assumption, stated as follows. Define $C_i(j) := 1$ if the $j$th attempt by node $i$ results in a collision, and $C_i(j) := 0$ if it results in a success.

**(A4)** [Bianchi's Decoupling Approximation] For each node $i \in \mathcal{N}$, the collision sequence $\{C_i(j)\}$ is i.i.d. with $P(C_i(j) = 1) = p_i$ for some constant $p_i$.

In reality successive attempts by the same node may occur if it repeatedly selects timer value 0 while other nodes' timers remain frozen. In such cases the above assumption ceases to hold. This phenomenon can be prominent when the window size is small, and has been taken into account in some recent work [15]. We call the string of successive attempts a *run of attempts*, and the first attempt in a run a *run-first-attempt* or simply *first-attempt*. If successive attempts were consider, we could alternatively adopt the following assumption: let $C'_i(j) := 1$ if the first-attempt of the $j$th run of attempts by node $i$ results in a collision, and $C'_i(j) := 0$ if it results in a success, and assume the first-attempt collision sequence (FACS) decoupling, i.e.,

**(A4')** [FACS Decoupling Approximation] For each node $i \in \mathcal{N}$, the first-attempt collision sequence $\{C'_i(j)\}$ is i.i.d. with $P(C'_i(j) = 1) = p_i$ for some constant $p_i$.

In this study we will ignore the possibility of successive attempts for simplicity of presentation and adopt **(A4)**. This is reasonable when the initial window size is sufficiently large. Our empirical results are fairly close between with and without consideration of successive attempts for large backoff windows. For small backoff windows, the discrepancy between the two will be illustrated in the numerical results.

We will use the term *backoff length* to mean the total number of slots that a node spends between two successive timer renewals during the service process, which is the selected timer value plus 1. Define $N_i^s$ and $N_i^{tx}$, respectively, as the numbers of slots and transmission attempts that node $i$ takes in serving one packet. $\overline{W}_i := \frac{\mathbb{E}[N_i^s]}{\mathbb{E}[N_i^{tx}]}$ is referred to as the average backoff length of node $i$.

Using Bianchi's approximation, we have

$$\begin{aligned}\mathbb{E}[N_i^s] &= \sum_{k=0}^{\infty}\sum_{j=0}^{k} \frac{2^{\min\{j,m\}}W + 1}{2}(p_i)^k(1-p_i)\\ &= \sum_{j=0}^{\infty} \frac{2^{\min\{j,m\}}W + 1}{2}\left(\sum_{k=j}^{\infty}(p_i)^k(1-p_i)\right)\\ &= \sum_{j=0}^{\infty} \frac{2^{\min\{j,m\}}W + 1}{2}(p_i)^j\end{aligned}$$

where $W$ is the size of the initial backoff window and $m$ is the value of the maximum backoff stage. Also note $\mathbb{E}[N_i^{tx}] = \frac{1}{1-p_i}$. Therefore, $\overline{W}_i$ is given by

$$\overline{W}_i = \frac{1}{2}\left[W\left((1-p_i)\sum_{j=0}^{m-1}(2p_i)^j + (2p_i)^m\right) + 1\right].$$

We next derive a relationship between the transmission attempt probability and $\hat{\rho}_i$. Let $\tau_i(n)$ be the probability that node $i$ initiates a transmission attempt in the $n$th slot.

*Lemma 1:* $\tau_i := \lim_{n\to\infty} \tau_i(n)$ exists and is given by $\tau_i = \hat{\rho}_i/\overline{W}_i$.

*Proof:* Let $T_X(n)$ denote the event that node $i$ initiates an attempt in the $n$th slot. Then

$$\begin{aligned}\tau_i(n) = P(T_X(n)|\hat{Q}_i(n) > 0)\cdot P(\hat{Q}_i(n) > 0) +\\ + P(T_X(n)|\hat{Q}_i(n) = 0)\cdot P(\hat{Q}_i(n) = 0).\end{aligned}$$

Consider now the sequence of slots in which node $i$ has a packet in service. Given the decoupling among nodes, the occurrences of slots in which node $i$ starts the service for a packet thus form renewal events. Regarding each transmission attempt as one-unit reward and using the renewal reward theory, we then obtain

$$\lim_{n\to\infty} P(T_X(n)|\hat{Q}_i(n) > 0) = \frac{\mathbb{E}[N_i^{tx}]}{\mathbb{E}[N_i^s]} = \frac{1}{\overline{W}_i}.^2$$

Since $\lim_{n\to\infty} P(\hat{Q}_i(n) > 0) = \hat{\rho}_i$, and $P(T_X(n)|\hat{Q}_i(n) = 0) = 0$, the result follows. ∎

To put the above result in context, one easily verifies that in the extreme case where all nodes are saturated and identical, we have $\hat{\rho}_i = \rho_i = \rho = 1$ and $p_i = p$ for all $i$. Consequently,

$$\begin{aligned}\tau_i = \tau &= \frac{2}{W\left((1-p)\sum_{j=0}^{m-1}(2p)^j + (2p)^m\right) + 1}\\ &= \frac{2(1-2p)}{(1-2p)(W+1) + pW(1-(2p)^m)},\end{aligned}$$

which is exactly the same as obtained in [7] Eqn (7).

## III. Single Channel Stability Region

### A. The stability region equation $\Sigma$

Our first main result is the following theorem on the quantitative description of $\Lambda$. Let $\mathbb{E}[S_{i,Q,\overline{T}_x}]$ denote the conditional average length of a slot given that the queue at node $i$ is non-empty but $i$ does not transmit in this slot. $T_s$ and $T_c$ denote the lengths of a successful transmission and a collision, respectively.

*Theorem 1:* $\boldsymbol{\lambda} \in \Lambda$ if and only if there exists at least one solution $\boldsymbol{\tau} = (\tau_1, \tau_2, \ldots, \tau_N)$ to the following constrained

[2] We note that [16] used a similar technique in computing the conditional transmission probability defined therein.

system of equations $(\Sigma, \mathrm{C}, \boldsymbol{\lambda})$:

$$\Sigma: \begin{cases} \tau_i = \dfrac{\hat{\rho}_i}{\overline{W}_i}, \forall i & \text{(a)} \\ p_i = 1 - \prod_{j \neq i}(1-\tau_j), \forall i & \text{(b)} \\ \rho_i = \min\left\{ \dfrac{\lambda_i}{P}\left( \dfrac{\overline{W}_i - 1}{1-p_i}\mathbb{E}[S_{i,Q,\overline{T}_x}] + \right.\right. & \\ \qquad \left.\left. + T_c \dfrac{p_i}{1-p_i} + T_s \right), 1 \right\}, \forall i & \text{(c)} \end{cases}$$

subject to

$$\mathrm{C}: \begin{cases} 0 \leq \tau_i \leq 1, \forall i & \text{(i)} \\ 0 \leq \rho_i < 1, \forall i & \text{(ii)} \end{cases}$$

where $P$ is the packet payload size.

*Proof:* $\Sigma$(a) is the result of Lemma 1, and $\Sigma$(b) is an immediate consequence of the definition of $p_i$. Let the average service time at node $i$ be $\overline{X}_i$ seconds per bit; the average service time per packet is thus $P\overline{X}_i$. Define $\overline{Y}_i(j)$ as

$$\overline{Y}_i(j) = T_c + \left( \frac{2^{\min\{j,m\}}W + 1}{2} - 1 \right)\mathbb{E}[S_{i,Q,\overline{T}_x}].$$

Physically, $\overline{Y}_i(j)$ is the average time between the beginning of the $j$th transmission attempt, which results in a collision, and the the beginning of the $(j+1)$th attempt, given that node $i$ encounters at least $j$ collisions before completing the service of some packet. Since the collision sequence is geometric, we have

$$\begin{aligned} P\overline{X}_i &= \sum_{k=0}^{\infty} \left[ \left( \frac{W+1}{2} - 1 \right)\mathbb{E}[S_{i,Q,\overline{T}_x}] + \sum_{j=1}^{k} \overline{Y}_i(j) + T_s \right] \times \\ &\quad \times (p_i)^k (1-p_i) \\ &= \sum_{j=1}^{\infty}\sum_{k=j}^{\infty} \overline{Y}_i(j)(p_i)^k(1-p_i) + \left( \frac{W+1}{2} - 1 \right) \times \\ &\quad \times \mathbb{E}[S_{i,Q,\overline{T}_x}] + T_s \\ &= \sum_{j=1}^{\infty} (p_i)^j \overline{Y}_i(j) + \left( \frac{W+1}{2} - 1 \right)\mathbb{E}[S_{i,Q,\overline{T}_x}] + T_s. \end{aligned}$$

Therefore,

$$\begin{aligned} P\overline{X}_i &= \sum_{j=1}^{\infty} \left[ (p_i)^j \left( T_c + \left( \frac{2^{\min\{j,m\}}W + 1}{2} - 1 \right) \times \right.\right. \\ &\quad \left.\left. \times \mathbb{E}[S_{i,Q,\overline{T}_x}] \right) \right] + \left( \frac{W+1}{2} - 1 \right)\mathbb{E}[S_{i,Q,\overline{T}_x}] + T_s \\ &= \sum_{j=0}^{\infty} \left[ \frac{2^{\min\{j,m\}}W - 1}{2} (p_i)^j \right] \mathbb{E}[S_{i,Q,\overline{T}_x}] + \\ &\quad + T_c \sum_{j=1}^{\infty} (p_i)^j + T_s \\ &= \frac{\overline{W}_i - 1}{1-p_i}\mathbb{E}[S_{i,Q,\overline{T}_x}] + T_c \frac{p_i}{1-p_i} + T_s. \end{aligned}$$

Note that $\tau_i < 1$ for all $i$, and we have $p_i < 1$ for all $i$ as a result. In addition, $\mathbb{E}[S_{i,Q,\overline{T}_x}]$ is finite (computed in the appendix). Hence we conclude that the packet service time is finite. Thus, the utilization factor of node $i$ is given by $\rho_i = \min\{\lambda_i \overline{X}_i, 1\}$ and $\Sigma$(c) follows. C(i) is for the validity of $\tau$ as a probability measure. ($\Sigma$, C(i), $\boldsymbol{\lambda}$) then constitutes a full description on the system utilization. C(ii) is the necessary and sufficient condition for stability as commented in the previous section. ■

For a given set of system parameter values, two sets of quantities are needed to compute $\Sigma$: $\mathbb{E}[S_{i,Q,\overline{T}_x}]$ and $\hat{\rho}_i$, $\forall i \in \mathcal{N}$. These are computed in Appendix B and C, respectively. In particular, in Appendix C we show that though it is analytically intractable, $\hat{\rho}_i$ is well approximated by

$$\hat{\rho}_i \approx \frac{\rho_i \mathbb{E}[S_{i,\overline{Q}}]}{\rho_i \mathbb{E}[S_{i,\overline{Q}}] + (1-\rho_i)\mathbb{E}[S_{i,Q}]},$$

where $\mathbb{E}[S_{i,Q}]$ (resp. $\mathbb{E}[S_{i,\overline{Q}}]$) is the conditional average length of a slot given that the queue at node $i$ is non-empty (resp. empty) at the beginning of this slot.

### *B. Characterizing the solutions to* $\Sigma$

Without the stability constraint C(ii), ($\Sigma$, C(i), $\boldsymbol{\lambda}$) can be rewritten as a vector equation in $[0,1]^N$, $\boldsymbol{\tau} = \boldsymbol{\Gamma}(\boldsymbol{\tau})$, where $\boldsymbol{\tau} = (\tau_1, \tau_2, \ldots, \tau_N) \in [0,1]^N$, and the existence of solutions can be shown by Brouwer's fixed point theorem. However, the uniqueness of its solution is in general difficult to prove; nevertheless, under the condition of a sufficiently large initial backoff window $W$, we have the following result on the uniqueness of its solution.

With a large initial backoff window $W$, the probability of collision is small, so we have $\overline{W}_i \approx \frac{W+1}{2}$. We also observe that $\mathbb{E}[S_{i,Q}] \approx \mathbb{E}[S_{i,\overline{Q}}]$ when $W$ is large (cf. Appendix B). Consequently, we can approximate $\hat{\rho}_i$ by $\rho_i$. Also, using the first-order Taylor approximation, we have $\prod_{j \neq i} \frac{1}{1-\tau_j} \approx 1 + \sum_{j \neq i} \tau_j$ for small $\tau$. Let $T_s = T_c = T$ for simplicity of presentation. Then $\Sigma$ can be approximated by the following system of equations,

$$\widetilde{\Sigma}: \begin{cases} \tau_i = \dfrac{\rho_i}{\frac{W+1}{2}} & \text{(a)} \\ \rho_i = \dfrac{\lambda_i}{P}\left[ \dfrac{W-1}{2}\left( \sigma + T\sum_{j \neq i} \tau_i \right) + T\left( 1 + \sum_{j \neq i} \tau_i \right) \right] & \text{(b)} \end{cases}$$

*Proposition 1:* $((\widetilde{\Sigma}), \boldsymbol{\lambda})$ admits a unique solution.

*Proof:* See Appendix E. ■

*Remark 2:* 1) The above result suggests that $\Sigma$ has a unique solution when $W$, the initial window size, is sufficient large. As an approximation we will take this condition to be equivalent to a large average backoff window. This is because the probability of a (first-attempt) collision decays inverse-linearly in $W$, and thus $\overline{W}_i$ is dominated by $W$ when $W$ is sufficiently large.

2) As we will see numerically in the next section, multiple fixed point solutions may arise when $W$ is small; this will be referred to as multi-equilibrium (as opposed to "multistable" or "metastable" [14] to avoid confusion).

In the proof of Proposition 1, we in fact obtained the approximated unique solution to $(\Sigma, \boldsymbol{\lambda})$. Therefore, by imposing feasibility constraints C, we can induce a simplified version of $(\Sigma, \mathrm{C}, \boldsymbol{\lambda})$ which is an approximation to $\Lambda$ and is easier to compute.

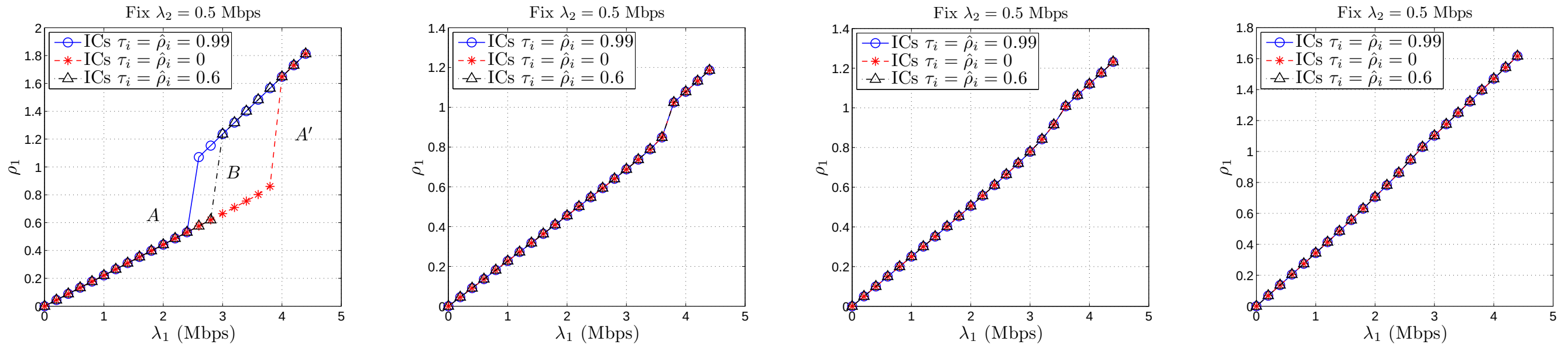

(a) $W = 2, m = 0$ (b) $W = 8, m = 5$ (c) $W = 32, m = 5$ (d) $W = 128, m = 5$

Fig. 1. Solution components for various scenarios: an illustration.

*Corollary 1:* When $W$ is sufficiently large, $\Lambda$ is approximated by

$$\tilde{\Lambda} = \left\{ \boldsymbol{\lambda} \in \mathbb{R}_+^N \;\middle|\; 0 < \frac{\gamma_i^1(\lambda_i) \sum_j \gamma_j^2(\lambda_i)}{1 - \sum_i \gamma_j^1(\lambda_i)} + \gamma_i^2(\lambda_i) < \frac{2}{W+1}, \forall i \right\},$$

where $\gamma_i^1(\lambda_i) = \frac{\lambda_i T}{P} \Big/ \left(1 + \frac{\lambda_i T}{P}\right)$, and $\gamma_i^2(\lambda_i) = \frac{\lambda_i((W-1)\sigma + 2T)}{P(W+1)} \Big/ \left(1 + \frac{\lambda_i T}{P}\right)$.

Within the context of a unique solution to $(\Sigma, \mathrm{C}, \boldsymbol{\lambda})$, consider $\boldsymbol{\lambda}$ as input parameters and rewrite $\Sigma$ as $\mathbf{F}(\boldsymbol{\tau}, \boldsymbol{\lambda}) = 0$, with $(n + n)$ unknowns ($\tau_i$'s and $\lambda_i$'s). We can then inspect the existence of an implicit function of $\boldsymbol{\tau}$ in terms of $\boldsymbol{\lambda}$, and for this we need to examine the invertibility of the corresponding Jacobian matrix. Note also that the correspondence between $\rho_i$ and $(\boldsymbol{\lambda}, \boldsymbol{\tau})$ given by $\Sigma$(c) is a continuous function. If the Jacobian is invertible on the boundary of the stability region $\Lambda$ in the space $\mathbb{R}_+^N$, then the continuity of $\rho_i = \rho_i(\boldsymbol{\lambda})$ is established. Hence, on the boundary of $\Lambda$, denoted by $\partial\Lambda$, there exists at least one node $i$ such that $\rho_i = 1$. Unfortunately, the invertibility of the Jacobian on $\partial\Lambda$ is highly non-trivial to determine and in general analytically intractable when the number of nodes is large. In the next section we numerically evaluate $(\Sigma, \mathrm{C}, \boldsymbol{\lambda})$ and more is discussed.

## IV. Numerical Results: Single Channel

Using $(\Sigma, \mathrm{C}, \boldsymbol{\lambda})$, we can quantitatively describe the stability region of a single channel system, and some numerical results for the two-user case are illustrated in this section. The parameters used in both the numerical computation and the simulation are reported in Table I in Appendix G. Under the basic access mechanism of DCF we have

$$\begin{cases} T_s = \frac{P}{\text{Tx. Rate}} + \text{Header} + \text{ACK} + \text{DIFS} + \text{SIFS} + 2\delta \\ T_c = \frac{P}{\text{Tx. Rate}} + \text{Header} + \text{DIFS} + \delta \end{cases}$$

where $\delta$ is the propagation delay.

### A. Multi-equilibrium and discontinuity in $\rho$

We first illustrate the existence of multi-equilibrium solutions and discontinuity of $\rho_i(\boldsymbol{\lambda})$ in $\boldsymbol{\lambda}$; this is shown in Figure 1. We fix the value of $\lambda_2$ and increase $\lambda_1$ from 0 to 4.5 Mbps. For each pair $\boldsymbol{\lambda} = (\lambda_1, \lambda_2)$, we solve for the fixed point(s) of $\boldsymbol{\Sigma}$ with the same set of initial values of $\tau_i$ and $\hat{\rho}_i$ for $i \in \mathcal{N}$ to which we refer as a set of initial conditions (ICs). We then convert the results to $\boldsymbol{\rho} = (\rho_i, i \in \mathcal{N})$ using Eqn. $\Sigma$(c). The collection of the pairs $(\boldsymbol{\lambda}, \boldsymbol{\rho}(\boldsymbol{\lambda}))$ then constitutes a *solution component* for this set of ICs. Note that this is obtained by solving ($\Sigma$, C(i), $\boldsymbol{\lambda}$) without considering the stability constraint C(ii). We repeat the above computation for different sets of ICs under the same system parameters including $W$ and $m$. The entire process is then repeated for different pairs ($W$, $m$). For each pair ($W$, $m$), the resulting solution components constitute an overall correspondence between the vectors $\boldsymbol{\lambda}$ and $\boldsymbol{\rho}(\boldsymbol{\lambda})$, and this is plotted for $\rho_1$ vs. $\lambda_1$ in Figure 1.

In the first scenario as shown in Figure 1(a), where the initial window is of the smallest possible size for two users and window expansion is disallowed ($m = 0$), three different zones of the correspondence $\rho_1(\lambda_1)$ are present, labeled as $A$, $A'$ and $B$ in the figure. In zones $A$ and $A'$, a single fixed point is admitted and $\rho_1(\lambda_1)$ reduces to a function, while in zone $B$ we see two solutions. Along each solution component, there is a jump in $\rho_1$ in zone $B$ as $\lambda_1$ increases; this is essentially a phase transition from stable to unstable regions. What this result illustrates is that depending on the initial condition, certain input rates may or may not lead to a feasible solution (a point in the stability region). Thus when such multi-equilibrium exists, we may have a collection of stability region $\Lambda$'s given different initial conditions, and this phenomenon is illustrated in Figure 3 and discussed in the next subsection in detail. Recall that under our definition of stability region and Theorem 1, an arrival rate vector is considered within the stability region as long as there exists such an initial condition that induces so; the stability region thus defined is therefore the supremum of this collection when multiple equilibria exist. The advantage of the "stability region" $A$ is that the points within are stable independent of the initial condition. With a slight abuse of terminology, we would later refer to this region as the stability region with multi-equilibrium.

Intuitively, initial conditions with large values suggest a pessimistic prediction on the system stability under $\boldsymbol{\lambda}$, and it may thus result in a small $\Lambda$; by contrast, ICs with small values render an optimistic one and a larger $\Lambda$. Empirically, we find that the set of ICs with $\tau_i = \rho_i \approx 1$ for $i \in \mathcal{N}$ results in the earliest jump in $\rho_1$ and the one with $\tau_i = \rho_i = 0$ for $i \in \mathcal{N}$ gives the latest. Consequently, solution components

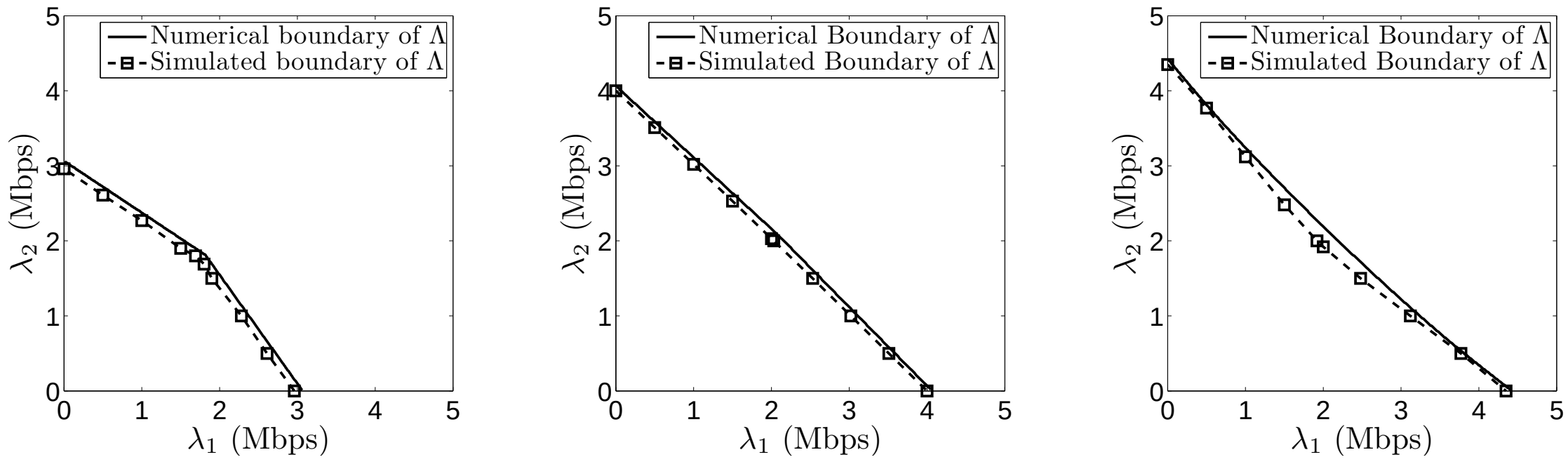

(a) $W = 128, m = 5$ (b) $W = 32, m = 5$ (c) $W = 8, m = 5$

Fig. 2. The stability regions in various scenarios - part I.

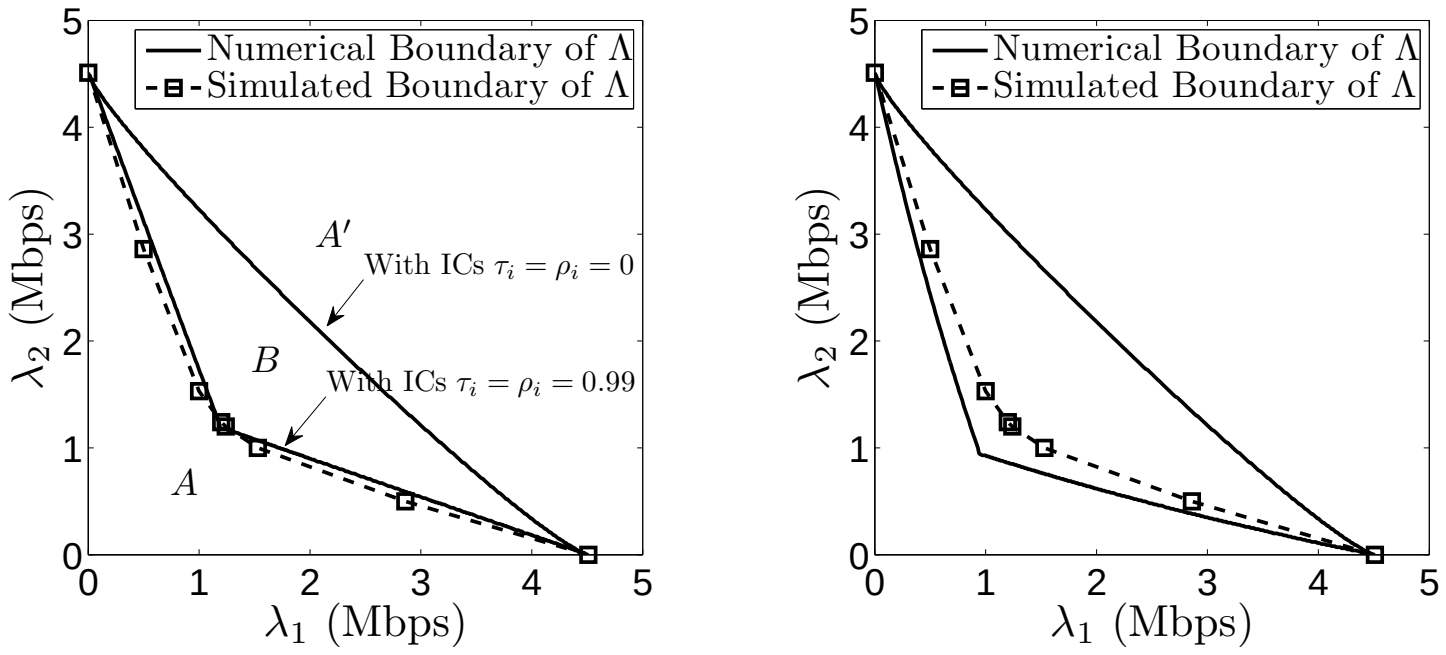

(a) Using Bianchi's decoupling approximation (b) Using the first-attempt decoupling approximation

Fig. 3. The stability regions in various scenarios - part II: $W = 2$ and $m = 0$.

resulting from these two sets of ICs define the boundary of zone $B$ and the corresponding stability regions, forming the empirical supremum and infimum of the collection of $\Lambda$'s.

Inspecting the set of figures Fig. 1(a)-1(d), we see that as the initial window increases, the multi-equilibrium gradually vanishes and the gap in $\rho_1$ caused by the jump discontinuity closes.

### B. Numerical and empirical stability regions

We numerically solve $(\Sigma, \mathrm{C}, \boldsymbol{\lambda})$ with two nodes to obtain the corresponding $\Lambda$, and then compare it with the simulated boundary. In simulation, for each fixed $\lambda_2$, we increase $\lambda_1$ with a step size $\Delta\lambda$, and compute the empirical throughput of node $i$ obtained under $\boldsymbol{\lambda}$, denoted as $S_i^{\boldsymbol{\lambda}}$, and the number of backlogged packets at node $i$ by the end of simulation, denoted as $B_i^{\boldsymbol{\lambda}}$. The simulator declares a point $\boldsymbol{\lambda}$ unstable if there exists at least one $i$ such that $S_i^{\boldsymbol{\lambda}} < \lambda_i$ and $B_i^{\boldsymbol{\lambda}} P/(\lambda_i T_f) > \alpha_i$, by the simulation time $T_f$, where $\alpha_i$ is an instability threshold, $0 < \alpha_i < 1$. In the experiment we set $\Delta\lambda = 0.1$ Mbps (100 Kbps), $T_f = 10$ sec and $\alpha_i = \alpha = 1\%$. The stable point $(\lambda_1, \lambda_2)$ such that $(\lambda_1 + \Delta\lambda, \lambda_2)$ is unstable is declared a point on the simulated boundary; the experiment is repeated for each $\lambda_2$ and the empirical mean value of $\lambda_1$ is recorded. Due to symmetry, only half of the boundary points are evaluated. The results are shown in Figure 2.

Our main observation is that when the initial (or average) backoff window is large, the stability region is convex (Figure 2(a)). The convexity gradually disappears as the window size decreases and the region is given by a near-linear boundary in Figure 2(b). It becomes clearly concave when the window size is small (Figure 2(c)). Interestingly, the case of $W = 32$ is the most frequently studied in the literature, and a linear boundary of the capacity region has been observed in [11]. As shown here, this linear boundary is only a special case in a spectrum of convex-concave boundaries. It is worth noting that in [12], Leith *et al.* established the general log-convexity of the rate region of 802.11 WLANs. This implies that the rate region could be either convex or concave, though [12] did not associate this with the window size as we have explicitly done here. It also suggests that the rate region and the stability region may be quite similar in nature; this however is not a formally proven statement, nor are we aware of such in the case of 802.11.

The change in the shape of the stability region as $W$ changes may be explained as follows. Small $W$ represents a highly aggressive configuration. This is much more beneficial when there is a high degree of asymmetry between the users' arrival rates. This is reflected in the concave shape of the region. When $W$ is large, users are non-aggressive, which is more beneficial when arrival rates are similar, resulting in the convex shape. Numerically, the $W = 8$ case gives the largest stability region. This seems to suggest that the largest stability region is given by the smallest choice of $W$ such that a unique feasible solution to $(\Sigma, \mathrm{C}, \boldsymbol{\lambda})$ exists. It would be very interesting to see if this could be established rigorously.

In Figure 3, we compute the stability regions of the case where $W = 2$ and $m = 0$ for two different sets of ICs. As discussed earlier, when multi-equilibrium exists we may have a collection of stability regions. This is clearly seen in Figure 3: three different zones $A$, $A'$ and $B$ in the correspondence $\rho_1(\lambda_1)$ are mapped accordingly onto $\Lambda$. From these results, we may interpret that in zones $A$ ($A'$), the system is uniformly stable (resp. unstable) regardless of the IC, while in zone $B$ the stability of system depends on the IC. As noted in [14], the simulated boundary reflects time-averages of multiple equilibria.

As mentioned earlier, for small backoff windows the occurrence of successive attempts is non-trivial, which our model has ignored. The first-attempt decoupling approximation **A4'** mentioned after **A4** captures the nodal behavior more accurately, and the adaptation of $\Sigma$ using this alternative assumption reduces to the computation of (conditional) mean slot length and is detailed in Appendix B. In Figure 3(b), we plot the counterpart of Figure 3(a) using the first-attempt decoupling approximation, and the discrepancy between results obtained using these two assumptions does exists. This is most notably shown in the numerical boundary $A$. The fact that the simulated boundary is now in between the two numerical boundaries verifies that this alternative assumption is more accurate. We do note however that for large windows this gap diminishes judging from numerical observation, which is to be expected.

### C. Discussion: from 802.11 DCF back to Aloha

We next recall results on the stability region of slotted Aloha, the natural prototype of modern 802.11 DCF, and provide an intuitive argument on why the qualitative properties of the stability region shown in the previous section are to be expected.

In [17], Massey and Mathys studied an information theoretical model of multiaccess channel which shares several fundamental features with slotted Aloha. They investigated the Shannon capacity region of this channel with $n$ users, which is shown to be the following subset of $\mathbb{R}_+^n$:

$$C = \left\{ \text{vect}\left( \tau_i \prod_{j \neq i} (1 - \tau_j) \right) \,\middle|\, 0 \le \tau_i \le 1, 1 \le i \le n \right\},$$

where $\text{vect}(v_i) = (v_1, v_2, \ldots, v_n)$, and $\tau_i$ is the transmission attempt rate of user $i$. In [13], Anantharam showed that the closure of the stability region of slotted Aloha is also given by $C$, under a geometrically distributed aggregate arrival process with parameter $1/(\sum_i \lambda_i)$ and probability $\lambda_i / \sum_j \lambda_j$ that such an arrival is at node $i$.

The above result on slotted Aloha can be used to explain the stability region of 802.11 DCF. Note that the main difference between the two lies in the collision avoidance mechanism. Instead of attempting transmission with probability $0 \le p \le 1$ in a slot under slotted Aloha, under DCF each user randomly chooses a backoff timer value within a window. The effect the average backoff length $\overline{W}$ has on transmission under DCF is akin to that of restricting the attempt rate $p$ within an upper bound $\frac{1}{\overline{W}}$ under slotted Aloha. Hence, the stability region of 802.11 DCF may be viewed as a subset of $C$ provided that we properly scale a slot to real time.

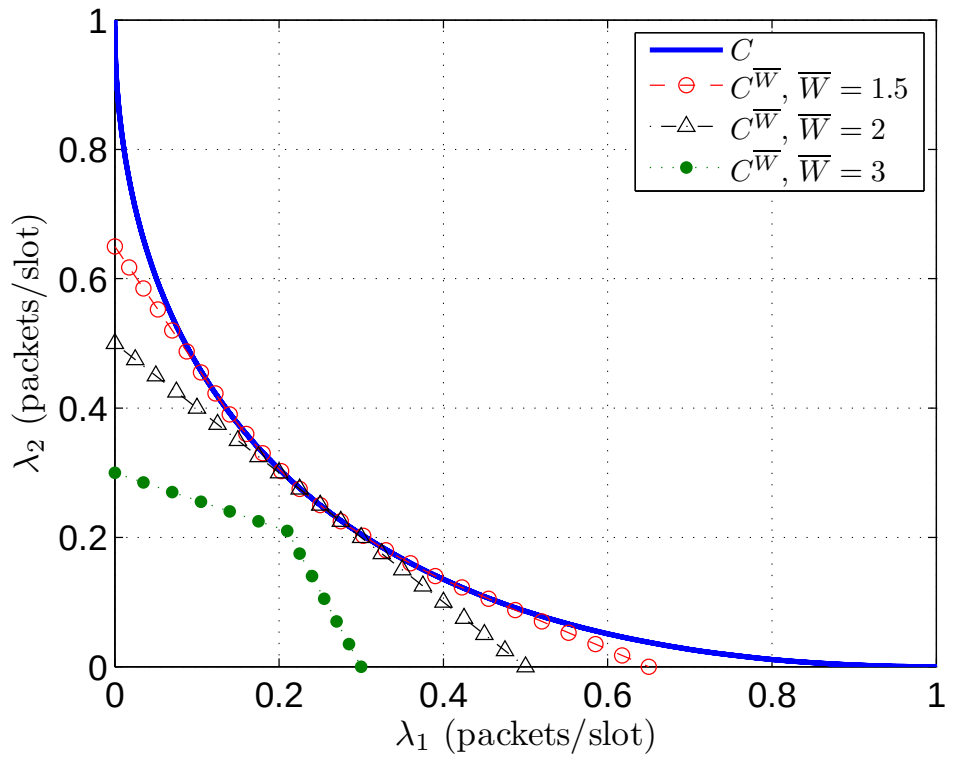

Fig. 4. The stability region of slotted ALOHA and induced subsets.

To verify this intuition, let $C^{\overline{W}}$ be the subset of $C$ when $0 \le p_i \le \frac{1}{\overline{W}}$ for all $i$. In Figure 4, we plot $C$ and $C^{\overline{W}}$ with different values of $\overline{W}$. We see that as $\overline{W}$ grows, $C^{\overline{W}}$ evolves from a concave set to a convex set, consistent with what we observed of 802.11 DCF in the previous subsection. It must be pointed out that this connection, while intuitive, is not a precise one technically. For instance, this connection might suggest that the stability region of 802.11 DCF will reduce to $C$ when the average backoff length is 1. This is however not true. In this trivial case, the stability region of 802.11 DCF is reduced to one dimension, i.e., the system is unstable for $n \ge 2$. This is because the retransmission probability of DCF is also lower bounded by the reciprocal of the window size at its backoff stage, and in the case when the backoff length is one another collision occurs with certainty.

## V. Multi-channel Analysis

Using a similar, mean-field Markovian model as we did in the single channel case, we can show that the stability region of a multi-channel system under a certain switching policy $\mathbf{g}$ is given by another system of equations denoted as $(\Sigma^{\mathbf{g}}, \mathrm{C}, \boldsymbol{\lambda})$, under the arrival rates $\boldsymbol{\lambda} = (\lambda_1, \lambda_2, \ldots, \lambda_N)$, and subject to the feasibility constraints C; this is given later in the section. In addition to the same set of assumptions made in the single channel model, we assume that the system has $K$ channels, indexed by the set $\mathcal{C} = \{1, 2, \ldots, K\}$.

The fundamental conceptual issue accompanying channelization is the notion of a channel switching policy, either centralized or distributed, that introduces channel occupancy and packet assignment distributions for each node. An additional technical issue induced by channelization is the heterogeneity of embedded time units among different channels. Since the slot length in a channel is by nature a random variable that depends on random packet arrivals, channels are in general strongly asynchronous in the embedded time units. Thus, as nodes switch among channels, we may need to switch the corresponding reference of embedded time in the slot based analysis. We therefore define the notion of a slot in different contexts as follows.

*Definition 3:* Consider the virtual backoff timer defined earlier separately for a *single channel*. A *channel-slot (c-slot)* is defined as the time interval between two consecutive decrements on this virtual timer for a given channel.

*Definition 4:* Consider a virtual backoff timer *at each node* that counts down indefinitely according to the node's backoff state, and is synchronized to the virtual timer of the channel in which the node resides and is done upon switching. A *node-slot (n-slot)* is defined as the time interval between two consecutive decrements on a given node's virtual backoff timer.

*Remark 3:* There is no inherent difference between the two types of slots. However, this differentiation of time references becomes crucial when we define quantities based on the random embedded time. This observation will be made more concrete in the analysis. We will also omit the explicit association of a channel (node) index with a slot whenever it does not cause ambiguity.

A channel switching or scheduling policy **g** induces a number of distributions related to $\Sigma^{\mathbf{g}}$. Denote by $\mathcal{Q}_i^n(j) = \{q_i^{(k)}(j), k \in \mathcal{C}\}$, where $q_i^{(k)}(j)$ is the probability that node $i$ is in channel $k$ at the beginning of its $j$th n-slot, $t_j^-$. $\mathcal{Q}_i^n(j)$ is referred to as the the channel occupancy distribution *in n-slots* of node $i$ in the $j$th n-slot.

Denote by $\mathcal{Q}_i^c(j) = \{\hat{q}_i^{(k)}(j), k \in \mathcal{C}\}$, where $\hat{q}_i^{(k)}(j)$ is the probability that node $i$ is in channel $k$ at the beginning of its $j$th c-slot, $\hat{t}_j^-$. $\mathcal{Q}_i^c(j)$ is referred to as the channel occupancy *profile* of node $i$ at the $j$th c-slot. Note that $\mathcal{Q}_i^c(j)$ is not necessarily a distribution and $\sum_{k\in\mathcal{C}} \hat{q}_i^{(k)}(j)$ need not be 1 for a given $j$.

Denote by $\mathcal{Q}_i^p(l) = \{\tilde{q}_i^{(k)}(l), k \in \mathcal{C}\}$, where $\tilde{q}_i^{(k)}(l)$ is the probability that the $l$th packet of node $i$ is served in channel $k$, and $\mathcal{Q}_i^p(l)$ is referred to as the packet assignment distribution of node $i$.

We have the following assumptions on policy **g**.

- **(A5)** Under **g**, Bianchi's approximation is still satisfied.
- **(A6)** **g** is independent of the binary state of the queue at any node (empty vs. non-empty).
- **(A7)** **g** is *nonpreemptive* in a channel for the entire service process of a packet; that is, a channel-switching decision is only made before or after the service process of a packet.
- **(A8)** The limits of $\mathcal{Q}_i^n(j)$, $\mathcal{Q}_i^c(j)$ and $\mathcal{Q}_i^p(l)$ exist under **g** as their respective arguments tend to infinity, and are denoted by $\mathcal{Q}_i^n$, $\mathcal{Q}_i^c$ and $\mathcal{Q}_i^p$, respectively.[3]

Similar as in single channel analysis, we impose the Markovian assumption on the discrete-time queueing process $\hat{Q}_i^{(k)}(n)$, which is the embedded process of $Q_i(t)$ (queue state of node $i$) sampled at the boundaries of c-slots of channel $k$, and define $\hat{\rho}_i^{(k)} = \lim_{n\to\infty} P(\hat{Q}_i^{(k)}(n) > 0)$. Also, let $\tau_i^{(k)}(n)$ be the probability that node $i$ initiates a transmission attempt in the $n$th c-slot of channel $k$. Then we have the following lemma; its proof is similar to that of Lemma 1 (based on **A6** and **A8**) and omitted.

*Lemma 2:* $\tau_i^{(k)} := \lim_{n\to\infty} \tau_i^{(k)}(n)$ exists and is given by $\tau_i^{(k)} = \hat{q}_i^{(k)} \hat{\rho}_i^{(k)} / \overline{W}_i^{(k)}$, where $\overline{W}_i^{(k)} := \frac{\mathbb{E}[N_i^{s,(k)}]}{\mathbb{E}[N_i^{tx,(k)}]}$ is the average backoff length of node $i$ in channel $k$, with $N_i^{s,(k)}$ and $N_i^{tx,(k)}$ defined in parallel as in the single channel case.

*Remark 4:* Under **A7**, $\overline{W}_i^{(k)}$ is given by

$$\overline{W}_i^{(k)} = \frac{1}{2}\left[W\left((1-p_i^{(k)})\sum_{j=0}^{m-1}(2p_i^{(k)})^j + (2p_i^{(k)})^m\right) + 1\right],$$

where $p_i^{(k)}$ is the probability of collision in channel $k$ given a transmission attempt and $W$ is the initial backoff window size.

Given any scheduling policy **g**, let $\Lambda^{\mathbf{g}}$ be the corresponding stability region, and we have the following theorem characterizing $\Lambda^{\mathbf{g}}$.

*Theorem 2:* $\boldsymbol{\lambda} \in \Lambda^{\mathbf{g}}$ if and only if there exists at least one solution $\boldsymbol{\tau} = (\boldsymbol{\tau}^{(k)}, k \in \mathcal{C})$ where $\boldsymbol{\tau}^{(k)} = (\tau_i^{(k)}, i \in \mathcal{N})$ to the following constrained system of equations $(\Sigma^{\mathbf{g}}, \mathrm{C}, \boldsymbol{\lambda})$,

$$\Sigma^{\mathbf{g}}: \begin{cases} \tau_i^{(k)} = \dfrac{\hat{q}_i^{(k)}\hat{\rho}_i^{(k)}}{\overline{W}_i^{(k)}}, \ \forall i,k & \text{(a)} \\ p_i^{(k)} = 1 - \prod_{j\neq i}(1-\tau_j^{(k)}), \ \forall i,k & \text{(b)} \\ \rho_i = \min\left\{\dfrac{\lambda_i}{P}\sum_{k\in\mathcal{C}}\left[\tilde{q}_i^{(k)}\left(\dfrac{\overline{W}_i^{(k)}-1}{1-p_i^{(k)}}\mathbb{E}[S_{i,Q,\overline{T}_x}^{(k)}] + \right.\right.\right. & \\ \quad \left.\left.\left. + T_c^{(k)}\dfrac{p_i^{(k)}}{1-p_i^{(k)}} + T_s^{(k)}\right)\right], 1\right\}, \ \forall i,k & \text{(c)} \end{cases}$$

subject to

$$\mathrm{C}: \begin{cases} 0 \le \tau_i^{(k)} \le 1, \ \forall i,k & \text{(i)} \\ 0 \le \rho_i < 1, \ \forall i & \text{(ii)} \end{cases}$$

where $i \in \mathcal{N}$ and $k \in \mathcal{C}$; $P$ is the packet payload size; $\mathbb{E}[S_{i,Q,\overline{T}_x}^{(k)}]$ is the conditional average length of a c-slot in channel $k$ given that the queue at node $i$ is non-empty but $i$ does not transmit in this slot.

*Proof:* The proof is an immediate extension of the proof of Theorem 1, given assumptions on **g**. ∎

The existence of a solution to $\Sigma^{\mathbf{g}}$ can be similarly established using Brouwer's fixed point theorem. We next study its uniqueness and the throughput optimality of a switching policy by resorting to an approximation given below, due to the complexity of $\Sigma^{\mathbf{g}}$. For the rest of this section, we will limit our discussion to the symmetric case where the channels have the same bandwidth and the system uses the same parameterization in all channels. We extend our discussion to more generic settings in the next section.

*Definition 5:* A scheduling policy is *unbiased* if the stationary channel occupancy distribution induced by such a policy is identical for every node, i.e., $q_i^{(k)} = q^{(k)}$ for all $i \in \mathcal{N}$ and $k \in \mathcal{C}$. It is denoted by $\mathbf{g}^U$, and the space of unbiased policies is $\mathcal{G}^U$.

[3]These limiting quantities are related by well-define correspondences, which are detailed in Appendix D, and those relations are used to numerically evaluate the stability region equation for a multi-channel system presented in this section.

We can obtain an approximation to $\Sigma^{\mathbf{g}^U}$ similarly as we did for $\Sigma$, using $\hat{q}^{(k)} \approx \tilde{q}^{(k)} \approx q^{(k)}$:

$$\widetilde{\Sigma}^{\mathbf{g}^U}: \begin{cases} \tau_i^{(k)} = \dfrac{q^{(k)}\rho_i}{\frac{W+1}{2}} & \text{(a)} \\ \rho_i = \dfrac{\lambda_i}{P}\displaystyle\sum_{k\in\mathcal{C}}\left\{q^{(k)}\left[\frac{W-1}{2}\left(\sigma + T\sum_{j\neq i}\tau_j^{(k)}\right)\right.\right. & \\ \qquad\left.\left. + T\left(1+\displaystyle\sum_{j\neq i}\tau_j^{(k)}\right)\right]\right\} & \text{(b)} \end{cases}$$

and we have the following result.

*Theorem 3:* Consider a system modeled by $\widetilde{\Sigma}^{\mathbf{g}^U}$ and the associated stability region $\Lambda^{\mathbf{g}^U}$. For all sufficiently large initial window sizes $W$, (i) the system of equations $(\Sigma^{\mathbf{g}^U}, \boldsymbol{\lambda})$ admits a unique solution, and (ii) $\mathbf{g}^U$ is throughput optimal within the class $\mathcal{G}^U$ if $q^{(k)} = \frac{1}{K}$ for all $k$. These are referred to as *equi-occupancy* policies.

*Proof:* We omit the proof on uniqueness, which is similar to the single-channel case; see Appendix F for the proof on throughput optimality. ■

The above results provide the following insights in addition to what we have observed in the single-channel case. Firstly, it's worth noting that $\Sigma^{\mathbf{g}}$ reduces to $\Sigma$ in the single-channel case by properly configuring related parameters, and $\Sigma^{\mathbf{g}}$ thus constitutes a unified framework in describing the stability region of 802.11 DCF.

Secondly, the uniqueness of the solution to $(\Sigma^{\mathbf{g}^U}, \boldsymbol{\lambda})$ is in fact true for even small windows. As an example, in Figure 5, we plot the numerical boundaries of stability regions for various window settings with equal channel occupancy. Compared to results in the single-channel case, convexity of the stability region is observed even with small backoff windows in the bi-channel case. Also, the numerical multi-equilibrium phenomenon disappears in this case. One way to explain this is by considering the discounting effect of channelization on the attempt rate. The attempt rate of each node in a channel is discounted by the occupancy probability in that channel. As discussed in the single-channel case, the attempt rate is roughly upper bounded by the reciprocal of the average backoff window size. Hence channelization has the effect of window expansion. The same explanation also applies to the observation that the stability region in a multi-channel system is nearly always convex.

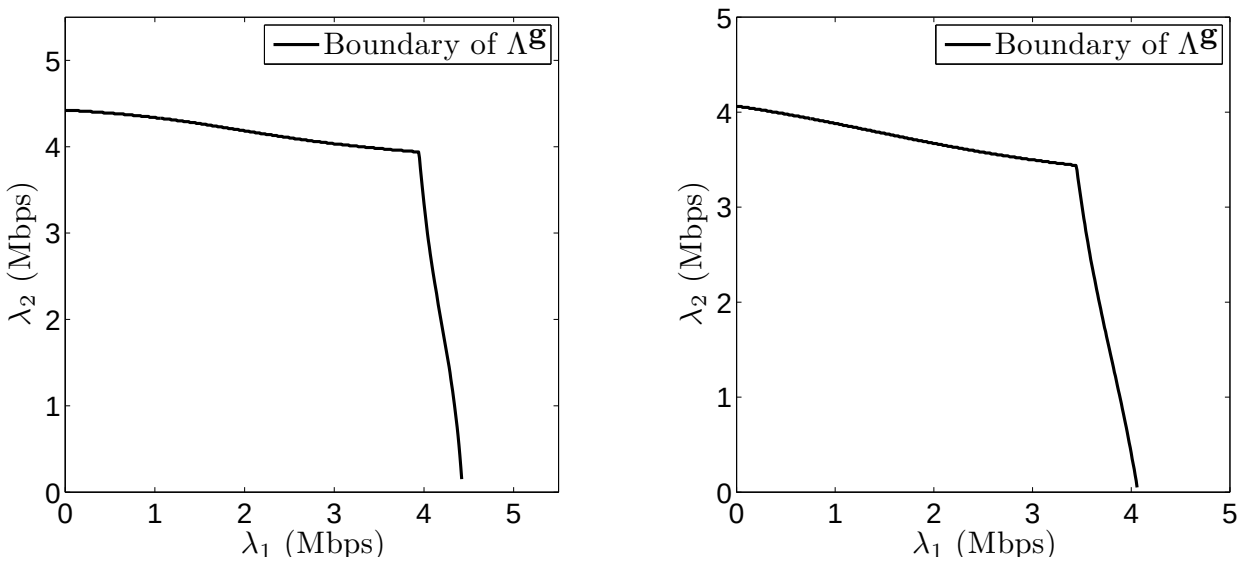

(a) $W = 8, m = 5$ (b) $W = 32, m = 5$

Fig. 5. The stability region of bi-channel 802.11 DCF under the equi-occupancy policy.

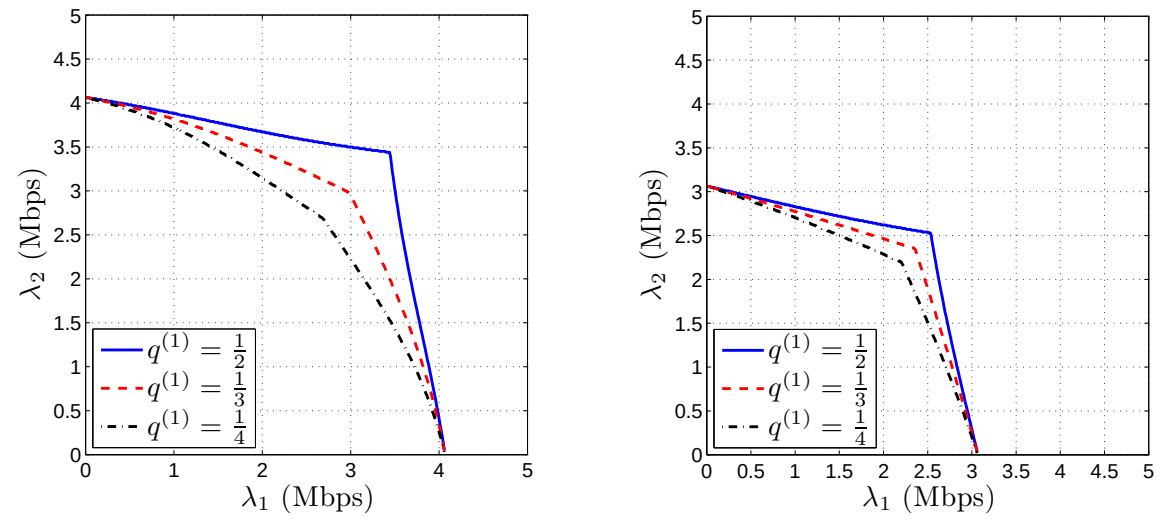

(a) $W = 32, m = 5$ (b) $W = 128, m = 5$

Fig. 6. Throughput optimality of equi-occupancy distribution.

Thirdly, given symmetric channelization, equal occupancy time is equivalent to equal packet assignment in each channel. The optimality of equi-occupancy policies therefore confirms the intuitive notion that load balancing (either in the number of active nodes or in the amount of date flow) optimizes the system performance in terms of expanding the stability region. In Figure 6, we plot the analytical boundaries of stability regions corresponding to different unbiased policies in two scenarios. As can been seen, the equi-occupancy policy results in a stability region that is the superset of those of the other unbiased policies. It is also worth noting that as the backoff window increases, the gap between the superset region and other inferior regions decreases, as the reciprocal of the window size becomes the dominant factor in upper bounding the attempt rate.

## VI. Applicability and Implementation of Unbiased Policies in Both Symmetric and Asymmetric Systems

In this section we discuss the applicability of the class of unbiased policies. We then present a number of practical implementations and their use in both symmetric and asymmetric systems.

### A. Unbiased policies

We have so far restricted our policy space to unbiased policies that induce a node-independent channel occupancy or packet assignment distribution. Note that while nodes in the same system are typically programmed with the same protocol stack, the same protocol may not necessarily yield the same statistical behavior among different nodes. Nevertheless, there are a number of circumstances in which node-independent behaviors are induced, which justifies our focus on unbiased policies. Firstly, if the protocol explicitly prescribes packet allocation to each channel, the resulting packet assignment distributions are identical for all nodes. Secondly, if nodes have identical arrival processes, they then have unbiased behavior as well. Unbiasedness can also be observed in a saturated network (however, such a system is unstable).

More generally, we note that when a node is active (i.e., its queue is non-empty and it is in the service process), from a mean-field point of view the channel conditions observed by this node is fully characterized by $p_i^{(k)}$ for each $k$ (as a result of the decoupling assumption), which is a function of $\tau_j^{(k)}$ for all $j \neq i$. Therefore, the set of attempt rates $\{\tau_i^{(k)}; \forall i, \forall k\}$ characterizes the contention condition in the system. If nodes are asymptotically symmetric, that is, $\lim_{N\to\infty} \tau_i^{(k)}/\tau_j^{(k)} = 1$, for all $i \neq j$ and $k$, then we have

$$\lim_{N\to\infty} \frac{p_i^{(k)}}{p_j^{(k)}} = \lim_{N\to\infty} \frac{1-\prod_{l\neq i}(1-\tau_l^{(k)})}{1-\prod_{l\neq j}(1-\tau_l^{(k)})}$$
$$= 1 + \lim_{N\to\infty} \frac{A(\tau_i^{(k)}-\tau_j^{(k)})}{A\tau_i^{(k)}+(1-A)} = 1,$$

where $A = \prod_{l\neq i,j}(1-\tau_l^{(k)})$. In this case we may consider the behavior induced by the underlying protocol on each node identical, and the corresponding policy unbiased. Note that the decoupling assumption is regarded as asymptotically true for a large number of nodes, so we may consider the asymptotic symmetry as an adjoint condition if we impose the decoupling approximation in modeling.

### *B. Practical implementation of throughput optimal unbiased policies: symmetric channels*

We have shown that when channels are symmetric the optimal switching policy within the class of unbiased policies is the equi-occupancy policy that balances load precisely. When channels are asymmetric, i.e., have different bandwidths, it is natural to expect that a load balancing policy yields throughput optimal performance, and to interpret a balanced load as having a packet assignment distribution proportional to the channel bandwidths. We will see that this interpretation is reasonable though not precise.

We begin by commenting on how such policies may be realized in a symmetric system.

We describe two very simple heuristics that implement an unbiased policy, and in particular, the equi-occupancy policy when channels are symmetric. The description is given in the bi-channel case for simplicity. The first is called SAS (switching after success), and the second SAC (switching after collision). In both schemes, a switching probability is assigned to each backoff stage. Under SAS (resp. SAC), a node switches to the other channel with probability $\alpha_i$ upon a successful transmission (resp. collision) if it is in the $i$th backoff stage when this success (resp. collision) occurs. In addition, in SAC, after switching to the other channel, a node does not reset its backoff stage; instead, it continues the exponential backoff due to the last collision. Note that SAS can be used to implement any arbitrary packet assignment distribution (and thus load distribution), which is a useful feature when we proceed to the implementation under asymmetric channels. This is because with the assumption of nonpreemptiveness of the policy, i.e., **A7**, switching after each successful transmission is equivalent to assigning packets.

These two schemes heuristically implement the equi-occupancy policy when channels are symmetric in the following sense. Consider the two-dimensional Markov chains for a bi-channel system in the form of Bianchi's model [7], where each state in one channel has a mirror state in the other. Since for both SAS and SAC, the corresponding Markov chain is irreducible with a finite number of states, using the argument of symmetry, the symmetric solution is the unique stationary distribution that reflects equi-occupancy. It should be noted however that neither of the above is a perfect solution and the key may be a proper combination of the two. The problem with SAS is that it can result in empty channels (the node that succeeded in the transmission happens to be the only node in that channel). When this happens nodes can tend to cluster in the non-empty channel for significant periods of time due to collision and backoff, while our mean field Markov analysis implicitly assumes no channels are empty for long. On the other hand, the problem with SAC (SAC rarely results in empty channels and avoids clustering in one channel) is that it interrupts the service process of a packet in a given channel, thus violating the nonpreemptive assumption about the policy.

### *C. Practical implementation of throughput optimal unbiased policies: asymmetric channels*

We next proceed to asymmetric channels and examine how these two heuristics perform in this setting, and in doing so also empirically examine when the stability region is maximized. In particular, we focus on the performance of a policy when the majority of the nodes have similar arrival rates, and we examine the advantage of load balancing in improving stability. In our experiment, we fix 10 nodes with arrival rate 0.5Mbps that creates a mean-field background in a bi-channel network while inspecting the stability region of another two nodes, which is the projection of the aggregate stability region onto a plane of these two nodes' arrival rates. All nodes use the same policy in a single experiment.

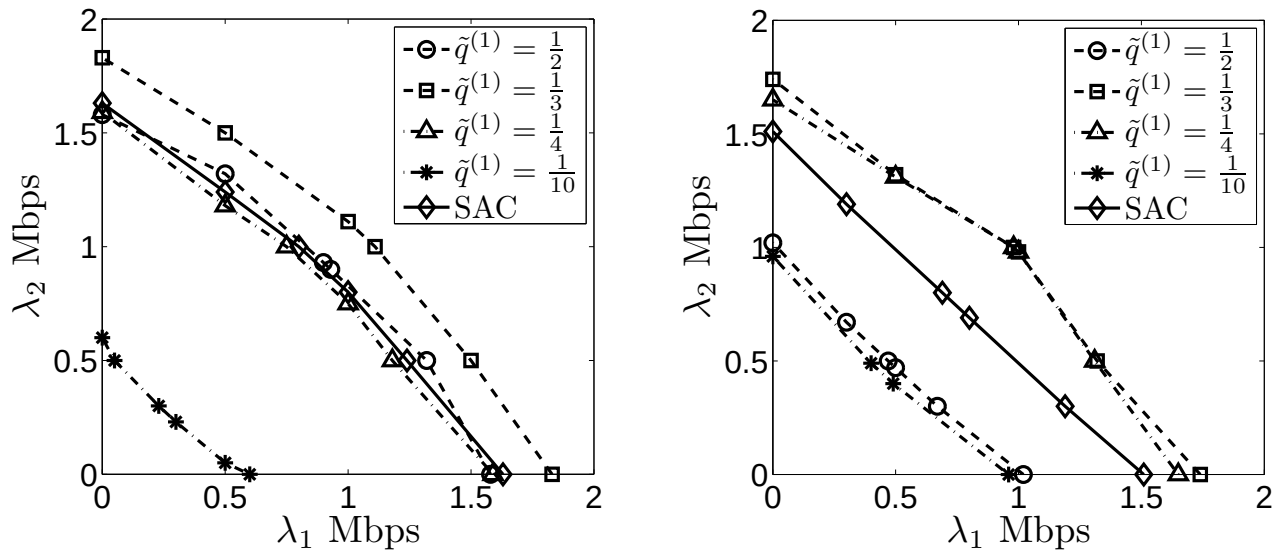

(a) The bandwidth ratio = 1:2 (b) The bandwidth ratio = 1:3

Fig. 7. The projection of simulated stability region onto a plane of arrival rates of the two nodes under inspection.

In Figure 7, we plot the empirical boundary of stability regions under different packet assignment distributions (implemented using SAS). As shown, policies with packet assignment ratio close to the bandwidth ratio indeed result in larger stability regions. However, while it seems safe to claim that properly balancing active time among channels according to their bandwidths improves the system performance, it remains unknown whether an exact match in load assignment is the optimal policy due to the nonlinearity of slot length in each channel w.r.t. active nodes. In addition, in practice we may not even know the effective bandwidth of each channel when channel conditions are imperfect.

It is therefore highly desirable to have an adaptive mechanism that dynamically adjusts the load distribution in practical implementation. Below we show that SAC to a large extent can achieve this goal, with the reason being that collision rate reflects the contention level and bandwidth information. Figure 7 also shows the empirical stability region obtained using SAC with switching probability at the $i$th backoff stage $\alpha_i = \frac{i}{m}$, where $m$ is the maximum backoff stage. SAC is clearly not optimal, but it maintains good performance under different bandwidth ratios.

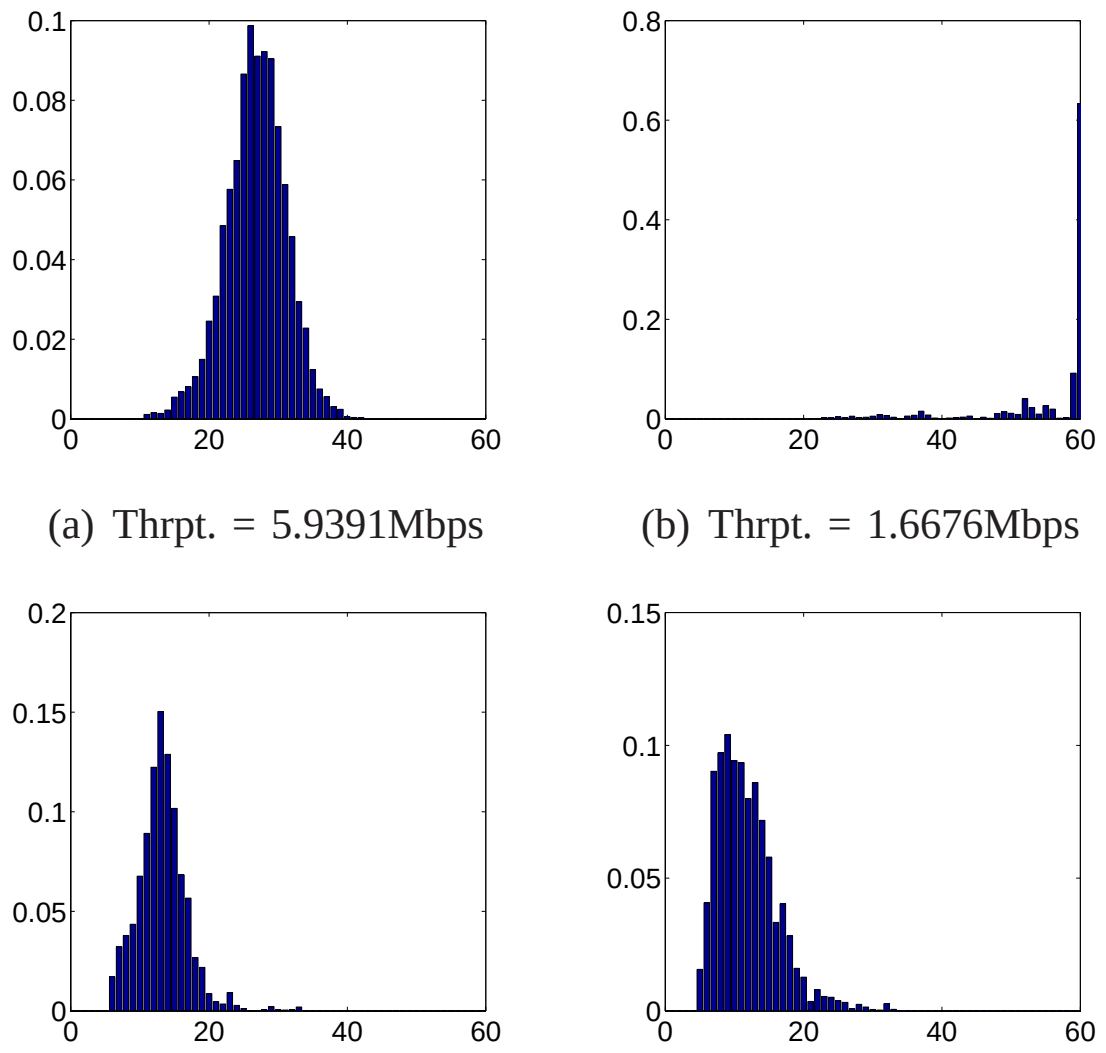

(a) Thrpt. = 5.9391Mbps (b) Thrpt. = 1.6676Mbps

(c) Thrpt. = 5.9720Mbps (d) Thrpt. = 5.9811Mbps

Fig. 8. Histogram of node population in the slower channel: (a)((b)) SAC (SAS) with $\alpha_i = 0.5$; (c)((d)) SAC (SAS) with $\alpha_i = \frac{i}{m}$.

We further highlight the adaptiveness of SAC in comparison to SAS. Assume that the active node population in each channel is the same and static, given then the same period of time, faster channels experience more transmission successes than slower ones. Therefore, if a SAS-like switching policy is adopted for a relatively congested network, nodes would cluster in the slower channels and the throughput performance degrades significantly. However, if the congestion is due to bandwidth asymmetry, then this is reflected in the collision rate of transmission, which in turns triggers channel reallocation under SAC. We illustrate this point using the following experiment. Consider a bi-channel system with strongly asymmetric channels, where the bandwidth of channel 1 (2) is 1Mbps (10Mbps). The system consists of 60 nodes each with an arrival rate 0.1Mbps, and this aggregate arrival rate (6Mbps) is slightly below the empirical saturation throughput under this setting. In the first test, we compare the resulting distribution of number of nodes in channel 1 between SAC and SAS with the switching probability $\alpha_i = 0.5$ for all stages in both channels, and we repeat the inspection with the switching probability $\alpha_i = \frac{i}{m}$ at stage $i$ in the second test; the duration of simulation is 180 seconds. The switching probability profile in the first test can be regarded as a blind configuration, while the second profile can be taken as an adaptive configuration that partially incorporates collision history into switching decisions. In Figure 8, we plot the histograms of the number of nodes in channel 1, as well as the empirical throughput obtained. As can be seen, the blindly configured SAS drives nodes to cluster in the slower channel, while SAC avoids this problem. Interestingly, SAS has comparable performance as SAC if we adjust the switching probabilities as we did in the second test, which reflects the congestion level in the residing channel, and both distributions "match" the bandwidth ratio. It suggests that while SAS is not as adaptive as SAC, it remains a valid alternative implementation and could achieve comparable performance when configured appropriately, as did above.

## VII. Conclusion

Using the characterization of the stability region of a multi-channel multi-user WLAN system, we investigated the throughput optimal channel switching schemes in such systems. This work can be extended in the following directions: 1) the effect of asymmetric channels on the characterization of stability region; 2) throughput optimal switching when considering the larger space of biased policies.

## Appendix A
## Proof of $\hat{\rho}_i \leq \rho_i$

We first define the following stochastic processes generated by the queueing process at node $i$.

$T_{i,Q}(t)/T_{i,\overline{Q}}(t) :=$ the total length of real time periods up to time $t$ that the queue at node $i$ is non-empty/empty (or $i$ is busy/idle);

$N_{i,Q}(t)/N_{i,\overline{Q}}(t) :=$ the total number of slots up to time $t$ that the queue at node $i$ is non-empty/empty at the beginning of slots.

These processes are well-defined on the same sample space $\Omega$. Assume that the queue is stable, then due to ergodicity $\rho_i$ and $\hat{\rho}_i$ can be expressed respectively as

$$\rho_i = \lim_{t\to\infty} \frac{T_{i,Q}(\omega,t)}{t} = \lim_{t\to\infty} \frac{T_{i,Q}(\omega,t)}{T_{i,Q}(\omega,t) + T_{i,\overline{Q}}(\omega,t)},$$

and

$$\hat{\rho}_i = \lim_{t\to\infty} \frac{N_{i,Q}(\omega,t)}{N_{i,Q}(\omega,t) + N_{i,\overline{Q}}(\omega,t)},$$

for all $\omega \in \Omega$. Let $\Delta_i(t)$ be the total time fragmentation of busy periods in idle slots of node $i$ up to time $t$, and let $S_{i,Q}(k)$ ($S_{i,\overline{Q}}(k)$) be the length of the $k$th busy (resp. idle) slot. Quantities described above are illustrated in Figure 9. Then, we have

$$T_{i,Q}(t) - \Delta_i(t) = \sum_{k=1}^{N_{i,Q}(t)} S_{i,Q}(k),$$

and

$$t = \sum_{k=1}^{N_{i,Q}(t)} S_{i,Q}(k) + \sum_{k=1}^{N_{i,\overline{Q}}(t)} S_{i,\overline{Q}}(k).$$

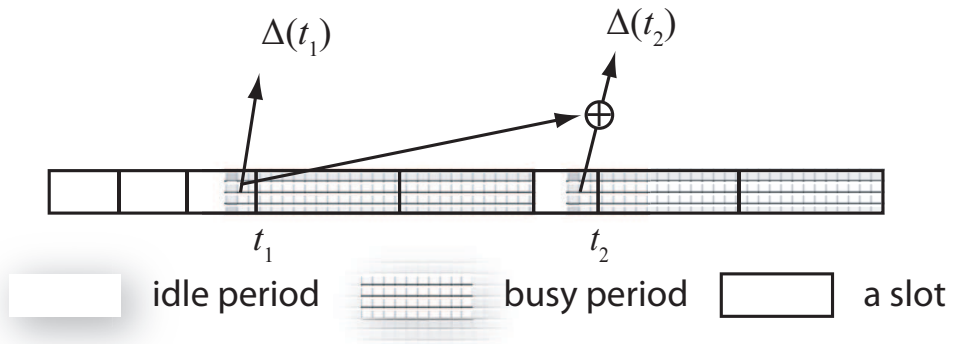

Fig. 9. Slotted time dynamics.

Therefore,

$$\rho_i \geq \lim_{t\to\infty} \frac{T_{i,Q}(t) - \Delta_i(t)}{t}$$
$$= \lim_{t\to\infty} \frac{\sum_{k=1}^{N_{i,Q}(t)} S_{i,Q}(k)}{\sum_{k=1}^{N_{i,Q}(t)} S_{i,Q}(k) + \sum_{k=1}^{N_{i,\overline{Q}}(t)} S_{i,\overline{Q}}(k)}$$
$$= \lim_{t\to\infty} \left[ \frac{\sum_{k=1}^{N_{i,Q}(t)} S_{i,Q}(k)}{N_{i,Q}(t)} N_{i,Q}(t) \Big/ \left( \frac{\sum_{k=1}^{N_{i,Q}(t)} S_{i,Q}(k)}{N_{i,Q}(t)} N_{i,Q}(t) + \frac{\sum_{k=1}^{N_{i,\overline{Q}}(t)} S_{i,\overline{Q}}(k)}{N_{i,\overline{Q}}(t)} N_{i,\overline{Q}}(t) \right) \right],$$

where we have suppressed the reference to a sample point $\omega$ in all involved processes for simplicity, or interpreted the equalities as with probability one. Let $\mathbb{E}[S_{i,Q}]$ and $\mathbb{E}[S_{i,\overline{Q}}]$ be the conditional average lengths of an arbitrary slot, given that the queue at node $i$ is non-empty or empty at the beginning of slot, respectively. We claim that $\mathbb{E}[S_{i,Q}] > \mathbb{E}[S_{i,\overline{Q}}]$ (see the next Appendix for justification). Note also that $N_{i,Q}(t) \to \infty$ and $N_{i,\overline{Q}}(t) \to \infty$ as $t \to \infty$ due to the stability assumption. Consequently, following ergodicity, we obtain

$$\rho_i \geq \lim_{t\to\infty} \frac{N_{i,Q}(t)\mathbb{E}[S_{i,Q}]}{N_{i,Q}(t)\mathbb{E}[S_{i,Q}] + N_{i,\overline{Q}}(t)\mathbb{E}[S_{i,\overline{Q}}]}$$
$$\geq \lim_{t\to\infty} \frac{N_{i,Q}(t)}{N_{i,Q}(t) + N_{i,\overline{Q}}(t)}$$
$$= \hat{\rho}_i.$$

When the queue is unstable, we have $\rho_i = \hat{\rho}_i = 1$. In either case, we have $\rho_i \geq \hat{\rho}_i$. It remains to justify the claim made above, which appears in the next Appendix.

## APPENDIX B
## COMPUTATION OF $\mathbb{E}[S_{\{\cdot\}}]$ AND RELATED QUANTITIES

Given an event $\{\cdot\}$, let $P_{idle;\{\cdot\}}$, $P_{succ;\{\cdot\}}$ and $P_{coll;\{\cdot\}}$ be the conditional probabilities that a slot is idle, that the transmission attempt in the slot is a success, and that the attempt is a collision, respectively. Notice that $P_{coll;\{\cdot\}} = 1 - P_{idle;\{\cdot\}} - P_{succ;\{\cdot\}}$. Therefore,

$$\mathbb{E}[S_{\{\cdot\}}] = \sigma \cdot P_{idle;\{\cdot\}} + T_s \cdot P_{succ;\{\cdot\}} + T_c \cdot P_{coll;\{\cdot\}}.$$

where $\sigma$, $T_s$ and $T_c$ are the lengths of an empty system slot, a successful transmission, and a collision, respectively. Define then by $\tau_{i,Q}$ the conditional probability that node $i$ transmits in an arbitrary slot, given its queue is non-empty at the beginning of this slot, and hence we have $\tau_{i,Q} = \frac{1}{W_i}$. Consequently,

$$P_{idle;i,\overline{Q}} = \prod_{j\neq i}(1-\tau_j),$$
$$P_{succ;i,\overline{Q}} = \sum_{j\neq i}\tau_j \prod_{l\neq i,j}(1-\tau_l),$$
$$P_{idle;i,Q} = (1-\tau_{i,Q})\prod_{j\neq i}(1-\tau_j),$$
$$P_{succ;i,Q} = \sum_{l}\tilde{\tau}_l \prod_{j\neq l}(1-\tilde{\tau}_l),$$

where

$$\tilde{\tau}_j = \begin{cases} \tau_{i,Q}, & \text{if } j = i \\ \tau_j, & \text{if } j \neq i \end{cases}.$$

Since $P_{idle;i,Q} < P_{idle;i,\overline{Q}}$ and $\sigma < \min\{T_s, T_c\}$, we have $\mathbb{E}[S_{i,Q}] > \mathbb{E}[S_{i,\overline{Q}}]$ and they are both finite. Explicit expressions for other variations of $\mathbb{E}[S_{\{\cdot\}}]$ can be derived in a similar way, and are thus omitted.

When successive attempts are considered and the FACS decoupling approximation is adopted, we can adapt the above computation as follows. Denote by $L_{idle;\{\cdot\}}$, $L_{succ;\{\cdot\}}$ and $L_{coll;\{\cdot\}}$ the average lengths of the slot in the corresponding cases, and hence $L_{idle;\{\cdot\}} = \sigma$, $L_{succ;\{\cdot\}} = T_s$ and $L_{coll;\{\cdot\}} = T_c$ in the above computation. When successive attempts are taken into account, we have

$$L_{idle;\{\cdot\}} = \sigma,$$
$$L_{succ;\{\cdot\}} = T_s \sum_{i=0}^{\infty}\left(\frac{1}{W}\right)^i = \frac{1}{1-\frac{1}{W}}T_s,$$

and

$$L_{coll;\{\cdot\}}$$
$$\approx T_c + \sum_{i=1}^{\infty}\left\{\left[\left(\frac{1}{\overline{CW}_{\{\cdot\}}}\right)^2\right]^i T_c + 2\left[\left(\frac{1}{\overline{CW}_{\{\cdot\}}}\right)^2\right]^{i-1} \times \times \frac{1}{\overline{CW}_{\{\cdot\}}}\left(1-\frac{1}{\overline{CW}_{\{\cdot\}}}\right)\frac{1}{1-\frac{1}{W}}T_s\right\}$$
$$= \frac{1}{1-\left(\frac{1}{\overline{CW}_{\{\cdot\}}}\right)^2}T_c + \frac{2}{\left(1+\overline{CW}_{\{\cdot\}}\right)\left(1-\frac{1}{W}\right)}T_s$$
$$\approx \frac{1}{1-\left(\frac{1}{W}\right)^2}T_c + \frac{2}{W-\frac{1}{W}}T_s,$$

where $\overline{CW}_{\{\cdot\}}$ is the conditional average backoff window size. These quantities are well-defined when $W \geq 2$ which is presumed in application. The first approximation of $L_{coll;\{\cdot\}}$ is due to omitting the possibility of collisions involving three or more nodes, and the other one results from substituting $\overline{CW}_{\{\cdot\}}$ with the initial backoff window size $W$. Note that, if we neglect successive attempts, we have $L_{succ;\{\cdot\}} = T_s$ and $L_{coll;\{\cdot\}} = T_c$, which is also a natural consequence when $W$ is sufficiently large in the above equations.

## APPENDIX C
## APPROXIMATION OF $\hat{\rho}_i$

Due to the analytical intractability of $\Delta_i(t)$, we are interested in proper approximations of $\hat{\rho}_i$ that can lead to good estimate of $\Lambda$; a good estimate in the context of stability study means a tight underestimation. Recall that $\hat{\rho}_i \leq \rho_i$ and equality holds if and only if $\rho_i = 1$ or $\rho_i = 0$; therefore by replacing $\hat{\rho}_i$ by $\rho_i$ in $\Sigma$(c), solutions to the resulting system of equations form an underestimation of $\Lambda$ but accurate when $\rho_i = 1$ or 0 for all $i$. Moreover, when $0 < \hat{\rho}_i < 1$, we have

$$\begin{aligned}\hat{\rho}_i &= \lim_{t\to\infty} \frac{\frac{T_{i,Q}(t)-\Delta_i(t)}{S^{\text{av}}_{i,Q}(t)}}{\frac{T_{i,Q}(t)-\Delta_i(t)}{S^{\text{av}}_{i,Q}(t)} + \frac{T_{i,\overline{Q}}(t)+\Delta_i(t)}{S^{\text{av}}_{i,\overline{Q}}(t)}} \\ &\le \lim_{t\to\infty} \frac{\frac{T_{i,Q}(t)}{T_{i,Q}(t)+T_{i,\overline{Q}}(t)} S^{\text{av}}_{i,\overline{Q}}(t)}{\frac{T_{i,Q}(t)}{T_{i,Q}(t)+T_{i,\overline{Q}}(t)} S^{\text{av}}_{i,\overline{Q}}(t) + \frac{T_{i,\overline{Q}}(\omega,t)}{T_{i,Q}(t)+T_{i,\overline{Q}}(t)} S^{\text{av}}_{i,Q}(t)} \\ &= \frac{\rho_i \mathbb{E}[S_{i,\overline{Q}}]}{\rho_i \mathbb{E}[S_{i,\overline{Q}}] + (1-\rho_i)\mathbb{E}[S_{i,Q}]} \\ &\le \rho_i,\end{aligned}$$

where

$$S^{\text{av}}_{i,Q}(t) = \frac{1}{N_{i,Q}(t)} \sum_{k=1}^{N_{i,Q}(t)} S_{i,Q}(k)$$

and defining

$$\hat{\hat{\rho}}_i = \frac{\rho_i \mathbb{E}[S_{i,\overline{Q}}]}{\rho_i \mathbb{E}[S_{i,\overline{Q}}] + (1-\rho_i)\mathbb{E}[S_{i,Q}]},$$

we have $\hat{\rho}_i \le \hat{\hat{\rho}}_i \le \rho_i$. Hence, substituting $\hat{\rho}_i$ with $\hat{\hat{\rho}}_i$ in $\Sigma$(c), we can obtain a tighter underestimation of $\Lambda$ than with $\rho_i$, thus trading off computational complexity for higher accuracy. Empirical results suggest that $\hat{\hat{\rho}}$ is sufficiently close to $\hat{\rho}$, and we use $\hat{\hat{\rho}}$ as $\hat{\rho}$ throughout our computation.

## Appendix D
## Computation of $\mathcal{Q}^c$ and $\mathcal{Q}^p$

We first define the following processes generated by the channel switching policy at node $i$.

$N^{(k)}_{+i}(t)/N^{(k)}_{-i}(t) :=$ the total number of c-slots in channel $k$ up to time $t$ that node $i$ is present at their beginning;

$S^{(k)}_{+i}(j)/S^{(k)}_{-i}(j) :=$ the length of the $j$th c-slot in channel $k$ given the presence (resp. absence) of node $i$ at its beginning;

$N_i(t) :=$ the total number of n-slots at node $i$ up to time $t$.

When a channel switching is scheduled, the edges of c-slots of the two channels that a node switches between may not be aligned. Hence, there may be a period of unsynchronized time of the nodal backoff timer, as shown in Figure 10. If omitting the unsynchronized time, we have

$$\sum_{l\in\mathcal{C}} \sum_{j=1}^{N^{(l)}_{+i}(t)} S^{(l)}_{+i}(j) = t = \sum_{j=1}^{N^{(k)}_{+i}(t)} S^{(k)}_{+i}(j) + \sum_{j=1}^{N^{(k)}_{-i}(t)} S^{(k)}_{-i}(j),$$

and then

$$\sum_{\substack{l\ne k\\ l\in\mathcal{C}}} \sum_{j=1}^{N^{(l)}_{+i}(t)} S^{(l)}_{+i}(j) = \sum_{j=1}^{N^{(k)}_{-i}(t)} S^{(k)}_{-i}(j),$$

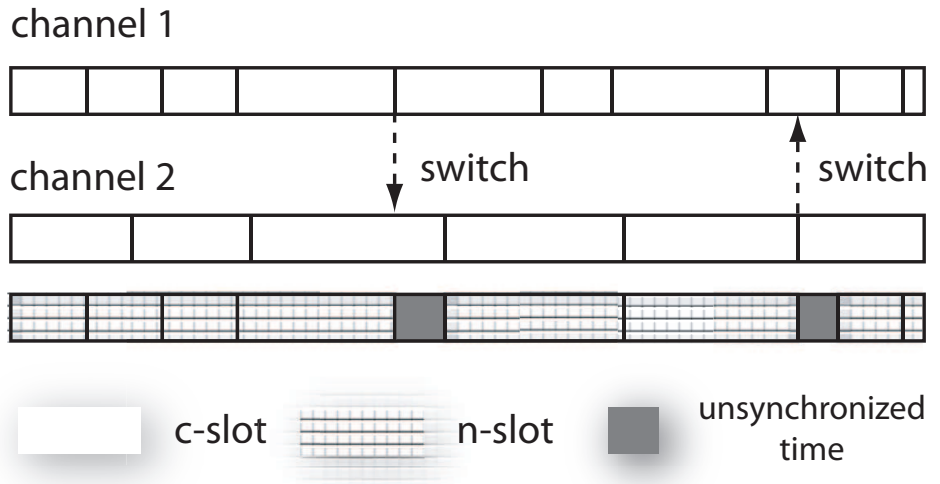

Fig. 10. Illustration of channel switching and timer synchronization.

or equivalently,

$$\sum_{\substack{l\ne k\\ l\in\mathcal{C}}} N^{(l)}_{+i}(t) \frac{\sum_{j=1}^{N^{(l)}_{+i}(t)} S^{(l)}_{+i}(j)}{N^{(l)}_{+i}(t)} = N^{(k)}_{-i}(t) \frac{\sum_{j=1}^{N^{(k)}_{-i}(t)} S^{(k)}_{-i}(j)}{N^{(k)}_{-i}(t)}.$$

Then $\hat{q}^{(k)}_i$ can be expressed alternatively as

$$\begin{aligned}\hat{q}^{(k)}_i &= \lim_{t\to\infty} \frac{N^{(k)}_{+i}(t)}{N^{(k)}_{+i}(t) + N^{(k)}_{-i}(t)} \\ &= \lim_{t\to\infty} \frac{\frac{N^{(k)}_{+i}(t)}{N_i(t)}}{\frac{N^{(k)}_{+i}(t)}{N_i(t)} + \sum_{\substack{l\ne k\\ l\in\mathcal{C}}} \frac{N^{(l)}_{+i}(t)}{N_i(t)} S^{\text{av},(l)}_{+i}(t) \Big/ S^{\text{av},(l)}_{-i}(t)} \\ &= \frac{q^{(k)}_i}{q^{(k)}_i + \sum_{\substack{l\ne k\\ l\in\mathcal{C}}} \left( \frac{q^{(l)}_i \mathbb{E}[slot^{(l)}_{+i}]}{\mathbb{E}[slot^{(k)}_{-i}]} \right)}.\end{aligned}$$

where

$$S^{\text{av},(l)}_{+i}(t) = \frac{\sum_{j=1}^{N^{(l)}_{+i}(t)} S^{(l)}_{+i}(j)}{N^{(l)}_{+i}(t)}.$$

We next define a few more generated processes at node $i$.

$P^{(k)}_i(t) :=$ the total number of packets transmitted by node $i$ in channel $k$ up to time $t$;

$B^{(k)}_i(j)/I^{(k)}_i(j) :=$ the number of busy slots in the service circle of the $j$th packet,

where a service circle is the period in slots between the beginning of service processes of two successive packets. A busy slot refers to a slot in the service process and a idle slot is a slot when the queue at the node is empty. Then, $q^{(k)}_i$ can also be expressed alternatively as

$$\begin{aligned}q^{(k)}_i &= \lim_{t\to\infty} \frac{N^{(k)}_{+i}(t)}{\sum_{l\in\mathcal{C}} N^{(l)}_{+i}(t)} \\ &= \lim_{t\to\infty} \frac{\sum_{j=1}^{P^{(k)}_i(t)} (B^{(k)}_i(j) + I^{(k)}_i(j))}{\sum_{l\in\mathcal{C}} \sum_{j=1}^{P^{(l)}_i(t)} (B^{(l)}_i(j) + I^{(l)}_i(j))} \\ &= \lim_{t\to\infty} \frac{M^{(k)}}{\sum_{l\in\mathcal{C}} M^{(l)}}.\end{aligned}$$

where

$$M^{(k)} = \frac{P^{(k)}_i(t)}{\sum_{h\in\mathcal{C}} P^{(h)}_i(t)} \times \frac{\sum_{j=1}^{P^{(k)}_i(t)} B^{(k)}_i(j)}{P^{(k)}_i(t)} \times$$

$$\times \frac{\sum_{j=1}^{P_i^{(k)}(t)} (B_i^{(k)}(j) + I_i^{(k)}(j))}{\sum_{j=1}^{P_i^{(k)}(t)} B_i^{(k)}(j)}.$$

We then obtain

$$q_i^{(k)} = \frac{\tilde{q}_i^{(k)} \overline{B}_i^{(k)} \frac{1}{\hat{\rho}_i^{(k)}}}{\sum_{l\in\mathcal{C}} \tilde{q}_i^{(l)} \overline{B}_i^{(l)} \frac{1}{\hat{\rho}_i^{(l)}}} = \frac{\tilde{q}_i^{(k)} \frac{\overline{W}_i^{(k)}}{1-p_i^{(k)}} \frac{1}{\hat{\rho}_i^{(k)}}}{\sum_{l\in\mathcal{C}} \tilde{q}_i^{(l)} \frac{\overline{W}_i^{(l)}}{1-p_i^{(l)}} \frac{1}{\hat{\rho}_i^{(l)}}}$$

$$= \frac{\tilde{q}_i^{(k)} \frac{\hat{q}_i^{(k)}}{\tau_i^{(k)}(1-p_i^{(k)})}}{\sum_{l\in\mathcal{C}} \tilde{q}_i^{(l)} \frac{\hat{q}_i^{(l)}}{\tau_i^{(l)}(1-p_i^{(l)})}},$$

where $\overline{B}_i^{(k)} := \frac{\sum_{j=1}^{P_i^{(k)}(t)} B_i^{(k)}(j)}{P_i^{(k)}(t)} = \mathbb{E}[N_i^{s,(k)}] = \frac{\overline{W}_i^{(k)}}{1-p_i^{(k)}}$.

## APPENDIX E
## PROOF OF PROPOSITION 1

Substituting $\widetilde{\Sigma}$(b) in (a), we obtain

$$\tau_i = \frac{2\lambda_i}{P(W+1)} \left[ \frac{W-1}{2} \left( \sigma + T \sum_{j\neq i} \tau_j \right) + T \left( 1 + \sum_{j\neq i} \tau_j \right) \right]$$

$$= \frac{2\lambda_i}{P(W+1)} \left[ \frac{W+1}{2} T \sum_{j\neq i} \tau_j + \frac{W-1}{2}\sigma + T \right]$$

$$= \frac{\lambda_i T}{P} \sum_{j\neq i} \tau_j + \frac{\lambda_i((W-1)\sigma + 2T)}{P(W+1)},$$

which can be rewritten as

$$\tau_i = \Big( \frac{\lambda_i T}{P} \sum_j \tau_j + \frac{\lambda_i((W-1)\sigma+2T)}{P(W+1)} \Big) \Big/ \Big( 1 + \frac{\lambda_i T}{P} \Big).$$

Therefore, let $y = \sum_j \tau_j$, $\gamma_i^1 = \frac{\lambda_i T}{P} \Big/ \Big(1 + \frac{\lambda_i T}{P}\Big)$ and $\gamma_i^2 = \frac{\lambda_i((W-1)\sigma+2T)}{P(W+1)} \Big/ \Big(1 + \frac{\lambda_i T}{P}\Big)$, and we have

$$\tau_i = \gamma_i^1 y + \gamma_i^2.$$

Then, $\widetilde{\Sigma}$ is equivalent to

$$\widetilde{\Sigma} : \begin{cases} \tau_i = \gamma_i^1 y + \gamma_i^2 & (a') \\ y = \sum_i \left( \gamma_i^1 y + \gamma_i^2 \right) & (b') \end{cases}$$

which admits only one solution, namely

$$\tau_i = \frac{\gamma_i^1 \sum_j \gamma_j^2}{1 - \sum_i \gamma_j^1} + \gamma_i^2.$$

## APPENDIX F
## PROOF OF THEOREM 3

Using $\widetilde{\Sigma}^{\mathbf{g}^U}$(a), we can rewrite $\widetilde{\Sigma}^{\mathbf{g}^U}$(b) as follows:

$$\rho_i = \frac{\lambda_i}{P} \sum_{k\in\mathcal{C}} \Big\{ q^{(k)} \Big[ \frac{W-1}{2} \Big( \sigma + T \sum_{j\neq i} \tau_j^{(k)} \Big) + T \Big( 1 + \sum_{j\neq i} \tau_j^{(k)} \Big) \Big] \Big\}$$

$$= \theta_i^1 \sum_{k\in\mathcal{C}} \Big( q^{(k)} \sum_{j\neq i} \tau_j^{(k)} \Big) + \theta_i^2$$

$$= \theta_i^1 \sum_{k\in\mathcal{C}} \phi_i(q^{(k)}; \rho_j, j\neq i) + \theta_i^2,$$

where $\theta_i^1 = \frac{\lambda_i(W+1)T}{2P}$, $\theta_i^2 = \frac{\lambda_i(W-1)\sigma+2T}{2P}$, and $\phi_i(q^{(k)};\rho_j, j\neq i) = q^{(k)} \sum_{j\neq i} \tau_j^{(k)} = \sum_{j\neq i} \alpha_j \left[q^{(k)}\right]^2$ with $\alpha_j = \frac{2\rho_j}{W+1} > 0$ for all $j$. Notice that $\phi_i(q^{(k)};\rho_j, j\neq i)$ is a convex function of $q^{(k)}$ given any fixed $\rho_j$ where $j\neq i$, and it is also an increasing function of $\rho_j$'s given any fixed $q^{(k)}$. We then have

$$\rho_i = \theta_i^1 \sum_{k\in\mathcal{C}} \phi_i(q^{(k)}) + \theta_i^2$$

$$= \theta_i^1 \cdot K \sum_{k\in\mathcal{C}} \left( \frac{1}{K} \phi_i(q^{(k)}) \right) + \theta_i^2$$

$$\geq \theta_i^1 \cdot K \phi_i \left( \sum_{k\in\mathcal{C}} \left( \frac{1}{K} q^{(k)} \right) \right) + \theta_i^2$$

$$= \theta_i^1 \cdot K \phi_i \left( \frac{1}{K} \right) + \theta_i^2,$$

where the equality holds when $q_i^{(k)} = \frac{1}{K}$. Therefore, when switching to the equi-occupancy policy from any arbitrary unbiased policy, the utilization factor of each node is always non-increasing. Hence, we conclude that the equi-occupancy scheduling policy is throughput optimal in $\mathcal{G}^U$.

## APPENDIX G
## MISCELLANEOUS

*Proof of the Markovian property of* $\hat{Q}_i(n)$

Let $\sigma_n$ denote the length of the $n$th slot, and let $\{P_t\}_{t\geq 0}$ be the transition semigroup of $Q_i(t)$. For any $q_i \in \mathbb{N}, i = 1, 2, \ldots, n$, we have

$$P(\hat{Q}_i(n+1) = q_{n+1} \mid \hat{Q}_i(n) = q_n, \ldots, \hat{Q}_i(0) = q_0)$$

$$= P\Big( Q_i(\sum_{j=1}^{n+1} \sigma_j) = q_{n+1} \mid Q_i(\sum_{j=1}^{n} \sigma_j) = q_n, \ldots, Q_i(0) = q_0 \Big)$$

$$= \int \cdots \int P\Big( Q_i(\sum_{j=1}^{n+1} t_j) = q_{n+1} \mid Q_i(\sum_{j=1}^{n} t_j) = q_n, \ldots,$$

$$Q_i(0) = q_0, \sigma_j = t_j, j = 1, \ldots, n+1 \Big) \times$$

$$\times f_{\sigma_{n+1},\ldots,\sigma_1}(t_{n+1}, \ldots, t_1 \mid Q_i(\sum_{j=1}^{n} \sigma_j) = q_n, \ldots, Q_i(0) = q_0)$$

$$dt_{n+1} \cdots dt_1$$

$$= \int \cdots \int P_{t_{n+1}}(q_n, q_{n+1}) f_{\sigma_{n+1}}(t_{n+1}) \times$$

$$\times f_{\sigma_n,\ldots,\sigma_1}(t_n, \ldots, t_1 \mid Q_i(\sum_{j=1}^{n} \sigma_j) = q_n, \ldots, Q_i(0) = q_0)$$

$$dt_{n+1} \cdots dt_1$$

$$= \int P_{t_{n+1}}(q_n, q_{n+1}) f_{\sigma_{n+1}}(t_{n+1}) \; dt_{n+1} \times$$

$$\times \int \cdots \int f_{\sigma_n,\ldots,\sigma_1}(t_n, \ldots, t_1 \mid Q_i(\sum_{j=1}^{n} \sigma_j) = q_n,$$

$$\ldots, Q_i(0) = q_0) \; dt_n \cdots dt_1$$

$$= \int P_{t_{n+1}}(q_n, q_{n+1}) f_{\sigma_{n+1}}(t_{n+1})\ dt_{n+1},$$

where the third equality is due to the assumption that $\sigma_i$'s are i.i.d. and independent of $Q_i(t)$, and $f$ is the (joint) probability density function of $\sigma_i$'s. On the other hand, for any $q_n, q_{n+1} \in \mathbb{N}$, we obtain

$$P(\hat{Q}_i(n+1) = q_{n+1} \mid \hat{Q}_i(n) = q_n)$$
$$= P\Big(Q_i(\sum_{j=1}^{n+1} \sigma_j) = q_{n+1} \mid Q_i(\sum_{j=1}^{n} \sigma_j) = q_n\Big)$$
$$= \int \cdots \int P\Big(Q_i(\sum_{j=1}^{n+1} t_j) = q_{n+1} \mid Q_i(\sum_{j=1}^{n} t_j) = q_n,$$
$$\sigma_j = t_j, j = 1, \ldots, n+1\Big) \times$$
$$\times f_{\sigma_{n+1},\ldots,\sigma_1}(t_{n+1}, \ldots, t_1 \mid Q_i(\sum_{j=1}^{n} \sigma_j) = q_n)\ dt_{n+1} \cdots dt_1$$
$$= \int \cdots \int P_{t_{n+1}}(q_n, q_{n+1}) f_{\sigma_{n+1}}(t_{n+1}) \times$$
$$\times f_{\sigma_n,\ldots,\sigma_1}(t_n, \ldots, t_1 \mid Q_i(\sum_{j=1}^{n} \sigma_j) = q_n)\ dt_{n+1} \cdots dt_1$$
$$= \int P_{t_{n+1}}(q_n, q_{n+1}) f_{\sigma_{n+1}}(t_{n+1})\ dt_{n+1} \times$$
$$\times \int \cdots \int f_{\sigma_n,\ldots,\sigma_1}(t_n, \ldots, t_1 \mid Q_i(\sum_{j=1}^{n} \sigma_j) = q_n)$$
$$dt_n \cdots dt_1$$
$$= \int P_{t_{n+1}}(q_n, q_{n+1}) f_{\sigma_{n+1}}(t_{n+1})\ dt_{n+1}.$$

Consequently, $Q_i(n)$ is a Markov chain. ∎

| | |
|---|---|
| Total bandwidth | 11 Mbps |
| Data packet length $P$ | 1500 Bytes |
| DIFS | 50 $\mu$s |
| SIFS | 10 $\mu$s |
| ACK packet length (in time units) | 203 $\mu$s |
| Header length (in time units) | 192 $\mu$s |
| Empty system slot time $\sigma$ | 20 $\mu$s |
| Propagation delay $\delta$ | 1 $\mu$s |
| Initial backoff window size $W$ | 32 |
| Maximum backoff stage $m$ | 5 |
| Data rate granularity $\Delta\lambda$ | 100 Kbps |
| Instability threshold constant | 1% |
| Total simulated time $T_f$ | 10 seconds |

TABLE I
SPECIFICATIONS OF THE IMPLEMENTATION OF TEST BENCH.